\documentclass[fleqn,usenatbib]{mnras}
\usepackage[section]{placeins}
\usepackage{newtxtext,newtxmath}
\usepackage{adjustbox}
\usepackage[T1]{fontenc}
\usepackage{ae,aecompl}

\usepackage{graphicx}	% Including figure files
\usepackage{amsmath}	% Advanced maths commands
\usepackage{multirow}
\usepackage{amsfonts}
\usepackage{mathtools}
\usepackage{algorithm, algpseudocode}
\usepackage{lineno}
\usepackage{hyperref}
\usepackage{xcolor}
\usepackage[english]{babel}
\usepackage{mleftright}
\usepackage{bm}
\usepackage{booktabs}
\usepackage{hyperref}

\newcommand{\dd}{\mathrm{d}}
\newcommand{\VS}{\mathbb{V}}
\newcommand{\condarg}[2]{\mleft(#1\middle\vert#2\mright)}
\renewcommand{\vec}{\mathbf}
\newcommand{\ncopt}{\texttt}

\title[APES: Approximate Posterior Ensemble Sampler]{APES: Approximate Posterior Ensemble Sampler}

\author[Vitenti, S.D.P., and Barroso E.J.]{
	Sandro D.~P.~Vitenti,$^{1}$\thanks{E-mail: vitenti@uel.br}
	Eduardo J.~Barroso,$^{1}$\thanks{E-mail: eduardo.jsbarroso@uel.br}
	\\
	$^{1}$Physics Department, Universidade Estadual de Londrina, Campus Universitário, CEP 86057-970, Londrina, Brasil \\
}

\date{Accepted 2023 July 18. Received 2023 June 30; in original form 2023 March 27}

\pubyear{2021}

\begin{document}
	\label{firstpage}
	\pagerange{\pageref{firstpage}--\pageref{lastpage}}
	\maketitle
	
	\begin{abstract}
		This paper proposes a novel approach to generate samples from target distributions that are difficult to sample from using Markov Chain Monte Carlo (MCMC) methods. Traditional MCMC algorithms often face slow convergence due to the difficulty in finding proposals that suit the problem at hand. To address this issue, the paper introduces the Approximate Posterior Ensemble Sampler (APES) algorithm, which employs kernel density estimation and radial basis interpolation to create an adaptive proposal, leading to fast convergence of the chains. The APES algorithm's scalability to higher dimensions makes it a practical solution for complex problems. The proposed method generates an approximate posterior probability that closely approximates the desired distribution and is easy to sample from, resulting in smaller autocorrelation times and a higher probability of acceptance by the chain. We compare the performance of the APES algorithm with the affine invariance ensemble sampler with the stretch move in various contexts, demonstrating the efficiency of the proposed method. For instance, on the Rosenbrock function, the APES presented an autocorrelation time 140 times smaller than the affine invariance ensemble sampler. The comparison showcases the effectiveness of the APES algorithm in generating samples from challenging distributions. This paper presents a practical solution to generating samples from complex distributions while addressing the challenge of finding suitable proposals. With new cosmological surveys set to deal with many new systematics, this method offers a practical solution for the upcoming era of cosmological analyses. The algorithms presented in this paper are available at \url{https://github.com/NumCosmo/NumCosmo}.
	\end{abstract}
	

	\section{Introduction}
	
	Markov Chain Monte Carlo (MCMC) methods have become a prominent tool in Bayesian inference since their inception by Metropolis in 1953~\citep{Metropolis1953}. MCMC has proven to be a versatile and effective method for generating samples from complicated and high-dimensional probability distributions, making it popular among scientists across various disciplines. In particular, MCMC methods have found widespread application in cosmology and astrophysics for estimating model parameters from observational data, as evidenced by several studies.
	
	Obtaining samples from the posterior distribution, which is computed using Bayes' theorem with prior distributions and the likelihood function of the data, can be computationally demanding, especially for high-dimensional problems~\citep{Trotta2008}. However, MCMC methods offer a viable solution to this issue by enabling the generation of samples from the posterior distribution, which can then be used to infer properties of the underlying physical process. The effectiveness of MCMC in this context has been instrumental in advancing our understanding of complex physical systems.
	
	Indeed, MCMC methods have played a significant role in addressing computational challenges in Bayesian inference, providing a powerful tool for generating samples from complex probability distributions. The ability to sample from the posterior distribution has enabled the exploration of parameter space and the estimation of uncertainties, making MCMC an essential tool in many scientific disciplines.
	
	MCMC methods have found numerous applications in cosmology, beginning with their early use in analyzing cosmic microwave data using Metropolis-Hastings algorithms~\citep{Christensen2001, Christensen2001a, Hastings1970}. In recent years, MCMC methods have been the subject of intense research in cosmology, with numerous studies exploring their use in various contexts. For instance, MCMC can be used to study the 21-cm signal, as done in~\cite{Greig2015} with \texttt{21cmMC} algorithm and in~\cite{Harker2012} to separate the 21-cm signal from foregrounds. The cosmic UV background has also been studied using MCMC methods~\citep{Caruso2019}. Furthermore, various works have focused on optimizing sampling techniques for cosmological parameter estimation~\citep{Dunkley2005}.
	
	While MCMC methods are a powerful tool for generating samples from posterior distributions, other efficient techniques exist, such as Nested Sampling (NS)~\citep{Skilling2006}. NS computes the integral of the posterior and produces a weighted sample of the posterior by generating a sample of the prior distribution. However, computing the marginal likelihood of the data required by NS can be computationally demanding, making the technique less efficient than MCMC methods in some cases. On the other hand, NS can be advantageous in performing Bayesian model comparison based on the Bayes factor~\citep{Congdon2006}. The effectiveness of NS can be seen in the \texttt{MultiNest} algorithm developed by~\cite{Feroz2009}, where the method was used to find constraints on the $\Lambda$CDM model, and in the \texttt{DYNESTY} algorithm~\citep{Speagle2020}, which has been applied to several real astronomical analyses~\citep{Leja2019, Zucker2019}. It should be noted that NS can be less efficient when there is a significant difference between the prior and posterior distributions of the parameters. Furthermore, the technique can require a significant amount of computational resources, especially for high-dimensional problems. For a recent review of NS and its applications, along with other NS-based samplers, we refer the reader to Ref.~\cite{Alsing2021} and the packages \texttt{Bibly-MCMC}~\citep{Ashton2021} and \texttt{POLYCORD}~\citep{Handley2015}.
	
	The development of ensemble samplers has led to the emergence of new algorithms that offer superior performance compared to traditional single-particle methods. For instance, Goodman's multiple-particle method~\citep{Goodman2010} is a notable improvement over such methods. One widely used ensemble sampler is \texttt{Emcee}, a Python implementation that employs parallel stretch moves and other techniques, making it a popular choice for Bayesian analysis in cosmology~\citep{Foreman-Mackey2013}. Similarly, the \texttt{CosmoHammer} package provides an optimized implementation of MCMC ensemble methods that is tailored to the requirements of the cosmology community for parameter estimation~\citep{Akeret2013}. The pairing of these methods with Boltzman codes as \texttt{CAMB} and \texttt{CLASS}~\citep{Lewis2000, Lesgourgues2011} has contributed to the popularity of MCMC in the cosmological and astronomy communities. Some packages that contain several of these codes are \texttt{COSMOMC}~\citep{Lewis2002}, \texttt{COSMOSIS}~\citep{Zuntz2015} and \texttt{MONTEPYTHON}~\citep{Audren2013}, which jointly have been used in numerous papers. 
	
	The ensemble sampler has been found to face significant challenges when applied to high-dimensional problems. For such, algorithms that exhibit superior performance for low-to-moderate dimensional problems often behave as a random walk sampler, leading to large self-correlation between points and high autocorrelation of the posterior sample. These issues compromise the efficiency of the algorithm and are commonly referred to as the curse of dimensionality~\citep{Jeffrey2020}. Further discussions about this problem can be found in~\cite{Morzfeld2019}.
	
	There have been several proposals to address the curse of dimensionality in high-dimensional problems, such as the ensemble slice sampling~\citep{Karamanis2020} and its python implementation in the \texttt{zeus} algorithm~\citep{Karamanis2021}, and the dynamic temperature selection~\citep{Vousden2016}. On another side, some ensemble sampler implementations tried to approximate or interpolate the likelihood to improve the code efficiency, such as \texttt{CMBFIT}~\citep{Sandvik2004}, \texttt{PICO}~\citep{Fendt2007}, \texttt{DASH}~\citep{Kaplinghat2002}, \texttt{SCoPE}~\citep{Das2014a} and others.  In this paper, we present a new method called the Approximate Posterior Ensemble Sampler (APES), which is a Metropolis-Hastings ensemble sampler that uses the APES move to generate high-dimensional samples with low autocorrelation and fast convergence. This proposal employs kernel density estimation and radial basis function interpolation to construct an approximation of the posterior, which is then used to propose points for the Metropolis-Hastings ensemble step. The result is an asymptotically high acceptance ratio and an effective solution to high-dimensional problems.
	
	The algorithms presented in this paper are implemented in the Numerical Cosmology library (NumCosmo)~\citep{Vitenti2012c}, which fully integrates the sampling algorithms with cosmological and astrophysical models.\footnote{The project repository can be found here: \href{https://github.com/NumCosmo/NumCosmo}{NumCosmo github}} The library is written in C and is developed in an object-oriented fashion using the GObject framework. It has automatic bindings for all languages that support GObject introspection, including Perl and Python and many others. Specifically, we utilize the Python interface to provide notebooks and examples demonstrating various applications of the algorithms.
	
	In \autoref{ensemblesampler}, we provide a detailed description of the ensemble sampler, including its key features and the stretch move that it employs. We use this as a basis for comparing our proposed sampler, which we introduce in \autoref{APES}. In this section, we provide a comprehensive explanation of the APES move, along with a thorough discussion of the techniques we use to implement it, including kernel density estimation and radial basis interpolation. To evaluate the effectiveness of our proposal, we present results in \autoref{comptests} for a range of test distributions, including the Rosenbrock distribution, a Gaussian Mixture distribution, the Funnel distribution, and a cosmology likelihood. We conclude with a discussion of the results and our main findings in \autoref{conclusion}.
	
	\section{Ensemble Sampler}
	\label{ensemblesampler}
	
	The main objective of the ensemble sampler is to obtain a sample from a probability distribution $\pi(x)$ defined on a $n$-dimensional parametric space $\VS \subset \mathbb{R}^n$. In our setting, this distribution is generally a posterior distribution computed from the likelihood of a dataset and a set of priors. Nevertheless, as usual, our treatment here applies to any probability distribution. One limitation of MCMC algorithms is that the transition kernel is conditional to the current position and cannot use more information about the sample to generate the next proposal point. In the literature, there are many different ways to circumvent this limitation. For example, a popular approach is the use of many parallel chains combined with an update method to improve the transition kernel, see~\cite{Lewis2002, Dunkley2005, Audren2013, Lewis2013}. More precisely, given the probability distribution $\pi(x)$ for $x \in \VS$ we define a Metropolis-Hastings transition kernel as
	\begin{equation}
		\label{KMH}
		\begin{split}
			T\condarg{y}{x} &= \alpha(x, y)q\condarg{y}{x}\\
			&+ \delta^n(x-y)\int_\VS \left[1-\alpha(x, z)\right]q\condarg{z}{x}\dd^nz,
		\end{split}
	\end{equation}
	where $y \in \VS$, $q\condarg{y}{x}$ is an arbitrary conditional probability distribution, which we call proposal distribution on $\VS$, $\dd^nz$ a measure on $\VS$, and the acceptance probability $\alpha(x, y)$ is defined by
	\begin{equation}\label{accept}
		\alpha(x, y) \equiv \mathrm{MIN}\left[1, \frac{q\condarg{x}{y}\pi(y)}{q\condarg{y}{x}\pi(x)}\right].
	\end{equation}
	
	It is easy to check that the above transition kernel satisfies the detailed balance condition, i.e., $T\condarg{y}{x}\pi(x) = T\condarg{x}{y}\pi(y)$. Consequently, the distribution $\pi(x)$ is left invariant by this kernel, that is,
	\begin{equation}\label{invariantpi}
		\int_\VS T\condarg{y}{x}\pi(x)\dd^nx = \pi(y).
	\end{equation}
	Furthermore, this is true for any intermediate step given by $q\condarg{y}{x}$. For this reason, one can provide any distribution $q\condarg{y}{x}$ (usually one that is easy to sample from) and build an algorithm that requires one evaluation of the distribution $\pi(x)$ per step. Of course, there are many caveats. A proposal distribution not well adapted to $\pi(x)$ (that is, frequently proposing points where $\pi(y)$ is very small) results in a tiny acceptance probability and consequently in repetitions on the chain and large auto-correlation. On the other hand, if $q\condarg{y}{x}$ frequently proposes points $y$ close to $x$, in the sense that $\pi(x) \approx \pi(y)$, they will have a high acceptance probability but this also results in chains with large auto-correlation.
	
	The parallel chain approach starts with a proposal distribution $q\condarg{y}{x}$. After some steps, one updates the proposal using the computed chains and restarts the algorithm. This method has some drawbacks. First, it is necessary to wait for the computation of some steps before updating the proposal. Second, one discards the computed points every time $q\condarg{y}{x}$ is updated. In this work, we use the ensemble sampler, which solves these problems by changing the probability space $\VS$ and $\pi(x)$, as we explain below.
	
	Instead of dealing with one $n$-dimensional point in the parametric space, the ensemble sampler considers a set of $L$ independent and identically distributed (iid) realizations of $\pi(x)$
	\begin{align}
		\label{ensemblepoint}
		\vec{x}&\equiv\left(x_1,x_2,\dots,x_L\right),
	\end{align}
	where $x_i \in \VS$ for $i = 1,2,\dots,L$. We refer to each component $X_i$ as a walker and an ensemble point as an $L \times n$ dimensional point $\vec{x} \in \mathbb{V}^{L}$. The target joint distribution for the iid realizations of $\pi(x)$ is
	\begin{align}
		\label{jointdist}
		\Pi(\vec{x})=\prod_{i=1}^{L}\pi(x_i).
	\end{align}
	As with other MCMC algorithms, the goal is to generate a sample from $\Pi(\vec{x})$. Since all points $x_i$ in $\vec{X}$ are iid, a sample of $N$ vectors $\vec{x}^j$, for $j=1,2,\dots,N$, produces a set of $N \times L$ realizations of $\pi(x)$.
	
	The ensemble sampler works as an MCMC algorithm, and thus the ensemble point $\vec{x}$ must be iterated until the chain reaches convergence. We label the steps (or time) using a superscript ${}^j$ starting at the $j = 0$, such that every time all components of $\vec{x}^j$ are updated we increase the time by one, which we call an ensemble sampler iteration. The advantage of this scheme is that the proposal distribution is conditioned on the previous ensemble point, that is, $Q\condarg{\vec{y}}{\vec{x}}$. For this reason, if $\vec{x}$ provides an approximated sample of $\pi(x)$ then it can be used to build a good proposal for the next step. Note that, in the conventional MCMC, the proposal $q\condarg{y}{x}$ can use a single point (which is usually taken as the location of $q\condarg{y}{x}$ to propose a new point), whereas now in an ensemble sampler the proposal can use a sample with $L$ elements to build the proposal distribution that is well adapted to $\pi(x)$ and consequently $\Pi(\vec{x})$. In this scenario, we avoid the two problems stated above since we can use a set of points to build the proposal without needing to restart the chains or discard any point.
	
	In our work, instead of building a proposal distribution $Q\condarg{\vec{y}}{\vec{x}}$ for the whole ensemble point $\vec{x}$, we use a strategy called \textit{partial resampling}, see for example~\cite{Liu2000, Liu2001}. In practice, this technique is similar to a Gibbs sampler. First, we define a partition of $\vec{x} = \left(x_i, \vec{x}_{[i]}\right)$ where $x_i \in \VS$ is the $i$-th component of $\vec{x}$ and $\vec{x}_{[i]}$ is an element of $\VS^{L-1}$ obtained by removing the $i$-th component of $\vec{x}$. Any transition kernel $T\condarg{y_i}{x_i,\vec{x}_{[i]}}$ that leaves invariant the conditional
	\begin{equation}\label{Picond}
		\Pi\condarg{x_i}{\vec{x}_{[i]}} = \frac{\Pi\mleft(\vec{x}\mright)}{\int_\VS\Pi\left(\vec{x}\right)\dd^n x_i} = \pi(x_i),
	\end{equation}
	also leaves $\Pi\mleft(\vec{x}\mright)$ invariant. In the last equality above, it is important to highlight that we made use of Eq.~\eqref{jointdist}, which indicates that the conditional probability on $\vec{x}_{[i]}$ in this context corresponds to the original target distribution. This is a well-known result that we reproduce here for completeness. Starting with the invariant condition on $\Pi\condarg{x_i}{\vec{x}_{[i]}}$, that is
	\begin{equation}
		\int_\VS T\condarg{y_i}{x_i,\vec{x}_{[i]}}\Pi\condarg{x_i}{\vec{x}_{[i]}}\dd^nx_i = \Pi\condarg{y_i}{\vec{x}_{[i]}},
	\end{equation}
	and substituting Eq.~\eqref{Picond}, we obtain
	\begin{equation}
		\int_\VS T\condarg{y_i}{x_i,\vec{x}_{[i]}}\Pi\mleft(\vec{x}\mright)\dd^nx_i = \Pi\mleft(y_i,\vec{x}_{[i]}\mright).
	\end{equation}
	
	In our case, the joint distribution is given by Eq.~\eqref{jointdist}, consequently, the proposal must leave the original distribution invariant, that is $\Pi\condarg{x_i}{\vec{x}_{[i]}} = \pi(x_i)$. Now, the advantage is that the proposal can depend on $\vec{x}_{[i]}$, i.e., it can be expressed by $q\condarg{y_i}{x_i,\vec{x}_{[i]}}$. Thus, the Metropolis-Hastings transition kernel~\eqref{KMH} can be used. This results in a kernel that not only leaves the distribution invariant but also satisfies the stronger requirement of the detailed balance assuring that the resulting chain is reversible, see for example~\cite{Robert2013}. In the current context, it is written as
	\begin{equation}
		\begin{split}
			T\condarg{y_i}{x_i,\vec{x}_{[i]}} &= \alpha_i(\vec{x}, y_i)q\condarg{y_i}{x_i,\vec{x}_{[i]}}\\
			&+ \delta^n(x_i-y_i)\int_\VS \left[1-\alpha_i(\vec{x}, z)\right]q\condarg{z}{\vec{x}}\dd^nz,
		\end{split}
	\end{equation}
	and the acceptance probability $\alpha_i(\vec{x}, y_i)$ for the $i$-th walker is defined by
	\begin{equation}
		\label{def:ap}
		\alpha_i(\vec{x}, y_i) \equiv \mathrm{MIN}\left[1, \frac{q\condarg{x_i}{y_i,\vec{x}_{[i]}}\pi(y_i)}{q\condarg{y_i}{\vec{x}}\pi\mleft(x_i\mright)}\right].
	\end{equation}
	Using this framework, we are allowed to use the remaining points $\vec{x}_{[i]}$ to build a proposal. As the chain converges, $\vec{x}_{[i]}$ will resemble a sample from $\pi(x)$, consequently by using the properties of $\vec{x}_{[i]}$ one can develop different proposals that are automatically updated by the properties of the sample and that keep the Markovian property of the chain.
	
	This method has already been used in the literature. For example~\cite{Goodman2010, Foreman-Mackey2013} developed a set of affine invariant proposals using the complementary set $\vec{x}_{[i]}$. The affine invariance property~\citep{Goodman2010, Coullon2020} assures that the sampler can efficiently sample from distributions that are not well distributed in all directions. The stretch move, for example, proved itself to be efficient for generic target distributions, but there are some issues when it comes to high-dimensional problems. As discussed in Ref.~\cite{Huijser2015}, the presence of a dimension-dependent term in the acceptance probability may cause its increase/decrease too fast in high-dimensional problems. Moreover, in high dimensions, there are spaces with low probability but with large volumes, and thus most of the proposals end up in those areas. This diminishes the acceptance ratio and increases the autocorrelation (since only proposals close to the originating point will be accepted) and can slow down the convergence.
	
	In this work, we introduce a new proposal for the ensemble sampler algorithm that uses kernel density estimation and radial basis interpolation to generate the proposal points for a Metropolis-Hastings algorithm. These methods are well suited for high-dimensional problems and thus can provide a robust algorithm for these settings.
	
	\section{The APES Proposal}
	\label{APES}
	
	In the ensemble sampler, the proposal distribution $q\condarg{y_i}{x_i,\vec{x}_{[i]}}$ may depend on the complementary set $\vec{x}_{[i]}$. The objective is to create an approximate distribution, $\tilde{\pi}\condarg{x}{\vec{x}_{[i]}}$, of the target $\pi(x)$ using the sample $\vec{x}_{[i]}$. However, in the \emph{partial resampling} scheme, it is necessary to update all components of $\vec{x}$ serially. The reason behind this is that once $x_i$ is updated to $x_i'$, the new ensemble point is given by $\vec{x}' = (x_1, x_2, \dots, x_i', \dots, x_L)$, and in order to update $x_{i+1}$, an approximation must be computed using $\vec{x}_{[i]}'$. This is problematic for two reasons: first, the need to recompute the approximation $L$ times for a single ensemble step, and second, it is not possible to parallelize the algorithm as each new proposal depends on the previously computed point. To address these issues, we propose splitting the ensemble point into two parts, a technique previously utilized in the \texttt{emcee}~\citep{Foreman-Mackey2013} and other similar algorithms. In this approach, $\vec{x} = (\vec{x}_{L_1}, \vec{x}_{L_2})$, where $\vec{x}_{L_1} = (x_1,x_2,\dots,x_{L/2})$ and $\vec{x}_{L_2} = (x_{L/2},x_{L/2+1},\dots,x_{L})$, and we assume $L$ is even for simplicity. We define two index sets as $L_1 = [1,2,\dots,L/2]$ and $L_2 = [L/2,L/2+1,\dots,L]$, and the function
	\begin{align}
		s(i) = \left\{\begin{array}{ccc}
			L_1 & \text{if} & i \in L_2 \\
			L_2 & \text{if} & i \in L_1
		\end{array}\right.
	\end{align}
	By exploiting the fact that $\vec{x}_{L_2}$ is a subset of $\vec{x}_{[i]}$ for any $i \in L_1$, we can build an approximation using $\vec{x}_{L_2}$ and use it to update all elements of $\vec{x}_{L_1}$ in parallel. Similarly, we can use the same approach to update $\vec{x}_{L_2}$ based on the approximation built from $\vec{x}_{L_1}$. This eliminates the need to recompute the approximation multiple times, which enables efficient parallelization. We now require a method to construct the approximation using $\vec{x}_{L_i}$ for $i=1,2$.
	
	Given a sample $\pmb{S}$ of $m$ points in $\mathbb{R}^n$, we apply the following strategy, first we use the natural vector space properties of $\mathbb{R}^n$ to define the Mahalanobis distance between $x\in\VS$ and a point $x_k \in \pmb{S}$ ($k=1,2,\dots,m$) as
	\begin{equation}\label{MDist}
		D^2_k(x, x_k) \equiv \mleft(x-x_k\mright)^{\intercal} \mathbf{C}_k^{-1}\mleft(x-x_k\mright), 
	\end{equation}
	where $\mathbf{C}_k$ is a positive definite symmetric $n \times n$ matrix. Then, the approximation is given by
	\begin{equation}\label{defapprox}
		\tilde{\pi}\condarg{x}{\pmb{S}} = \sum_{k =1}^m\frac{w_k}{h^n}K_k\mleft[\frac{D_k(x, x_k)}{h}\mright],\qquad \sum_{k =1}^m w_k = 1,
	\end{equation}
	where $h$ is the bandwidth parameter, $w_k\geq0$ the weights and $K_k$ a normalized multivariate distribution on $x$, that is,
	\begin{equation}
		K_k\mleft[D_k(x, x_k)\mright] \geq 0, \forall x\in\mathbb{R}^n, \quad \int_{\mathbb{R}^n}K_k\mleft[D_k(x, x_k)\mright]\dd^n x = 1.
	\end{equation}
	Generating a sample from $\tilde{\pi}\condarg{x}{\pmb{S}}$ is computationally straightforward. The process consists in selecting a kernel at random, based on the weights, and generating one random realization from that kernel. To use the approximation in Eq.~\eqref{defapprox}, the following must be defined: matrices $\mathbf{C}_k$ and distances $D^2_k(x, x_k)$ (\autoref{sec:dist}), kernel function $K$ (\autoref{sec:K}), weights $w_k$ (\autoref{sec:interpopt}), and bandwidth $h$ (\autoref{sec:bandwidth}). The methods for determining these quantities are discussed in \autoref{sec:apespopt}.
	
	Given the approximation method, we set the proposal distribution to be the approximation itself, i.e., $q\condarg{y_i}{x_i,\vec{x}_{[i]}} = \tilde{\pi}\condarg{x}{\vec{x}_{s(i)}}$. The acceptance probability, defined in Eq.~\eqref{def:ap}, is then given by:
	\begin{equation}\label{defalpha}
		\alpha_i\mleft(\vec{x}_{s(i)}, y_i\mright) = \mathrm{MIN}\left[1, \frac{\tilde{\pi}\condarg{x_i}{\vec{x}_{s(i)}}\pi(y_i)}{\tilde{\pi}\condarg{y_i}{\vec{x}_{s(i)}}\pi\mleft(x_i\mright)}\right].
	\end{equation}
	If $\tilde{\pi}\condarg{x}{\vec{x}_{s(i)}}$ is a good approximation of $\pi(x)$, such that $\tilde{\pi}\condarg{x}{\vec{x}_{s(i)}} = \pi(x)\left[1 + \delta(x)\right]$ with $\vert\delta(x)\vert \ll 1$, then we have:
	\begin{equation}
		\alpha_i\mleft(\vec{x}_{s(i)}, y_i\mright) = \mathrm{MIN}\left[1,\frac{1+\delta(x_i)}{1+\delta(y_i)}\right] \approx 1+\mathrm{MIN}\left[0,\delta(x_i)-\delta(y_i)\right].
	\end{equation}
	As a result, even for relatively large errors, of up to $\delta(x) \sim 20\%$, the acceptance probability may remain larger than that of many other MCMC samplers, and it is close to one when the errors are small. The following section will provide a detailed explanation of the step proposal and the algorithm being proposed in this paper.
	
	\subsection{The APES Move}
	\label{apesmove}
	
	As in any MCMC algorithm, we need a starting point $\vec{x}^0$. The main difference is that now $\vec{x}^0$ is a sample of points in $\VS^L$. Naturally, we cannot start with a sample from $\pi(x)$. Two options for obtaining $\vec{x}^0$ include sampling from the priors or computing the best-fit $x_\mathrm{bf}$ and Fisher matrix of $\pi(x)$ and sampling from the Gaussian approximation of $\pi(x)$. Both options are discussed in more detail in the next section and are available in our implementation.
	
	The algorithm starts with the determination of an initial point $\vec{x}^0$. The approximation $\tilde{\pi}\condarg{x}{\vec{x}^0_{L_2}}$ is then calculated, from which $L/2$ samples are drawn, resulting in $\vec{x}^{\star1}_{L_1}$. The acceptance probability for each sample $x_k^{\star1}$ is determined using Eq.~\eqref{defalpha}, which is given by:
	\begin{equation}\label{acceptprob}
		\alpha_k\mleft(\vec{x}_{L_2}^0, x_k^{\star1}\mright) = \mathrm{MIN}\left[1, \frac{\tilde{\pi}\condarg{x^0_k}{\vec{x}_{L_2}^0}\pi\mleft(x_k^{\star1}\mright)}{\tilde{\pi}\condarg{x_k^{\star1}}{\vec{x}_{L_2}^0}\pi\mleft(x_k^0\mright)}\right],
	\end{equation}
	forming the tuple $(t=1)$:
	\begin{equation}
		\pmb{\alpha}^t_{L_1} = \left[\alpha_k\mleft(\vec{x}_{L_2}^{t-1}, x_k^{\star t}\mright)\right]_{k \in L_1}.
	\end{equation}
	If the sample $x_k^{\star1}$ is accepted, $x_k^{1} = x_k^{\star1}$, otherwise $x_k^{1} = x_k^{0}$.  The algorithm then proceeds to calculate the next set of points, $\vec{x}_{L_2}^1$. This is done by computing the approximation $\tilde{\pi}\condarg{x}{\vec{x}^1_{L_1}}$ and drawing $L/2$ samples, referred to as $\vec{x}^{\star1}_{L_2}$. These samples are then accepted with a probability given by:
	\begin{equation}
		\alpha_k\mleft(\vec{x}_{L_1}^1, x_k^{\star1}\mright) = \mathrm{MIN}\left[1, \frac{\tilde{\pi}\condarg{x^0_k}{\vec{x}_{L_1}^1}\pi\mleft(x_k^{\star1}\mright)}{\tilde{\pi}\condarg{x_k^{\star1}}{\vec{x}_{L_1}^1}\pi\mleft(x_k^0\mright)}\right],
	\end{equation}
	which compose the second tuple $(t=1)$:
	\begin{equation*}
		\pmb{\alpha}^t_{L_2} = \left[\alpha_k\mleft(\vec{x}_{L_1}^{t}, x_k^{\star t}\mright)\right]_{k \in L_2}.
	\end{equation*}
	The new point $x_k^1$ is then set to $x_k^{\star1}$ if the probability is accepted, otherwise it remains $x_k^0$. After these two steps the next sample point $\vec{x}^1$ is determined. Now, these steps are repeated until the desired time $t$ is reached. The full step is illustrated in Algorithm~\ref{alg:walker_apes}.
	\begin{algorithm}[t]
		\caption{APES Algorithm}
		\label{alg:walker_apes}
		\begin{algorithmic}[1]
			\State Separate the $L$ walkers into two blocks $L_1$ and $L_2$
			\State Compute the initial sample with $L$ points $\vec{x}^0$ and set $t=0$
			\While{$t \leq t_\mathrm{max}$}
			\State Compute $\tilde{\pi}\condarg{x}{\vec{x}^t_{L_2}}$ and draw $\vec{x}^{\star t+1}_{L_1}$
			\For{$k \in L_1$}{ (parallelized loop)}
			\State Compute $\pi\mleft(x^{\star t+1}_{k}\mright)$ and $\alpha_k\mleft(\vec{x}_{L_2}^{t},x_k^{\star t+1}\mright)$
			\EndFor
			\State Update $\vec{x}_{L_1}^{t+1}$ using $\pmb{\alpha}^{t+1}_{L_1}$ computed above
			\State Compute $\tilde{\pi}\condarg{x}{\vec{x}^{t+1}_{L_1}}$ and draw $\vec{x}^{\star t+1}_{L_2}$
			\For{$k \in L_2$}{ (parallelized loop)}
			\State Compute $\pi\mleft(x^{\star t+1}_{k}\mright)$ and $\alpha_k\mleft(\vec{x}_{L_1}^{t+1},x_k^{\star t+1}\mright)$
			\EndFor
			\State Update $\vec{x}_{L_2}^{t+1}$ using $\pmb{\alpha}^{t+1}_{L_2}$ computed above
			\State Set $\vec{x}^{t+1} = \left(\vec{x}_{L_1}^{t+1},\;\vec{x}_{L_2}^{t+1}\right)$
			\State $t = t+1$
			\EndWhile
		\end{algorithmic}
	\end{algorithm}
	
	\subsection{APES Options}
	\label{sec:apespopt}
	
	The APES proposal's success relies on the accuracy of approximations made using two sub-samples, $L_i$, at each iteration. This accuracy is connected to the choice of distance function and its associated covariance matrix, $\mathbf{C}_k$, the kernel function $K_k$, as well as the bandwidth parameter, $h$, and the weights, $w_k$, for $k \in L_i$. The implementation offers two options for distances, a "same kernel approach" that uses a single distance function and a "variable kernel approach" that employs a different distance for each kernel. The weights can be calculated through either kernel density estimation or Radial Basis Function (RBF) interpolation. Finally, the bandwidth $h$ can be determined through either a Role of Thumb (RoT) or cross-validation. In the next subsections, we discuss these options in detail, providing a more thorough understanding of the implementation.
	
	\subsubsection{Distance functions}
	\label{sec:dist}
	
	In the same kernel approach, a single covariance matrix, $\widehat{\mathbf{C}}\left(\vec{x}_{L_i}\right)$, is used for all distance computations, meaning $\mathbf{C}_k = \widehat{\mathbf{C}}\left(\vec{x}_{L_i}\right)$. This matrix can be computed using an unbiased estimator based on the sample $\vec{x}_{L_i}$, as shown in the equation below:
	\begin{equation}
		\widehat{C}_\mathrm{b}\mleft(\vec{x}_{L_i}\mright) = \frac{1}{L/2-1}\sum_{k \in L_i} \left(x_{k} - \bar{x}\right)\left(x_k - \bar{x}\right)^\intercal, \; \bar{x} \equiv \frac{2}{L}\sum_{k \in L_i} x_{k}.
	\end{equation}
	In our study, we recognized the potential presence of outliers in the sample $\vec{x}_{L_i}$, particularly when the sample size is small or the chain has not yet converged. To mitigate this issue, we incorporated an Orthogonalized Gnanadesikan-Kettenring (OGK) algorithm~\citep{Maronna2002} into our implementation, which enables robust estimation of the covariance matrices $\widehat{C}_\mathrm{OGK}\mleft(\vec{x}_{L_i}\mright)$. To implement the OGK algorithm, we utilized the NumCosmo tools, which are built on top of the Lapack and BLAS library~\citep{Angerson1990, Lawson1979}. We estimated the underlying robust scale using the $Q_n$ robust scale estimators, which are implemented in the GSL library~\citep{GSLPC2010}. By incorporating the OGK algorithm into our implementation, we were able to better account for potential outliers in the sample and produce more robust covariance estimates for downstream analyses.
	
	In the variable kernel approach, instead of using a single covariance matrix for all computations, a different covariance matrix $\mathbf{C}_k$ is determined for each $x_k$ with $k\in L_i$ using a defined number of the nearest points to $x_k$ from $\vec{x}_{L_i}$. This approach allows for greater adaptability, but it may not be the best choice for cases with a low number of walkers in the ensemble. The difficulty in finding the nearest points of a position in multi-dimensional problems arises from differing scales between dimensions. For instance, in a 2-dimensional scenario where $x = (\theta_1, \theta_2)$ with variances $\mathrm{Var}(\theta_1) = 1$ and $\mathrm{Var}(\theta_2) = 10^4$ and the points $x_1 = (-3, 30)$, $x_2 = (3, 30)$ and $x_3 = (-3, 40)$. Using an identity matrix to calculate the distance $D^2(x, x')$ shows that $D(x_1, x_2) < D(x_1, x_3)$, even though $x_1$ and $x_2$ differ by six standard deviations in $\theta_1$ while $x_1$ and $x_3$ are within one standard deviation in both dimensions. To address this, we compute $\widehat{\mathbf{C}}\left(\vec{x}_{L_i}\right)$ following the same process as in the single kernel approach and using one of the available options for estimating covariance. Then, we use this matrix to compute the distances between points in $\vec{x}_{L_i}$.
	
	We define the number of nearest neighbors $m_k$ of $x_k$ as a fraction $p$ of $L/2$, rounded up to the nearest integer, i.e., $m_k = \lceil p L/2 \rceil$. We then use a fast k-Nearest Neighbor (kNN) search algorithm implementation~\citep{Ma2017} to find the $m_k$ closest points to $x_k$ defined as the subset $\vec{x}_{L_i}[\vec{x}_k, p] \subset \vec{x}_{L_i}$ using the distance calculated from the estimated global covariance matrix $\widehat{\mathbf{C}}\left(\vec{x}_{L_i}\right)$. Finally, we compute the covariance matrix $\mathbf{C}_k$ of each of these sub-samples $\vec{x}_{L_i}[\vec{x}_k, p]$ for use in the kernel $K_k$.  It is worth noting that different algorithms can be used to estimate the covariance matrix as discussed above.
	
	The variable kernel approach offers versatility in various scenarios. It proves effective in modeling multi-modal distributions $\pi(x)$ by adapting to the variance at each point and providing a more accurate approximation. Additionally, it handles distributions where the correlations between variables in $\VS$ vary significantly from point to point. The only drawback is that it requires a larger number of walkers, which is proportional to the dimension of $\VS$.
	
	\subsubsection{Kernel Function}
	\label{sec:K}
	
	In order to efficiently compute the approximation in Eq.~\eqref{defapprox} and facilitate easy sampling, a suitable kernel function $K_k$ must be selected. This study proposes two options: Gaussian (Gauss) and Student's t (ST) kernels. Both of these kernels are widely recognized and have readily accessible efficient implementations. Their mathematical expressions are respectively:
	\begin{align}
		\label{kg}K^\mathrm{G}_k(d) &= \frac{\exp \left[-\frac{d^2}{2}\right]}{\sqrt{(2 \pi)^{n}\mleft|\mathbf{C}_k\mright|}}, \\
		\label{kt}K^\mathrm{ST}_k(d) &= \frac{\Gamma[(\nu+n) / 2]}{\Gamma(\nu / 2) \sqrt{\left(\nu\pi\right)^{ d}\left|\mathbf{C}_k\right|}}\left[1+\frac{d^2}{\nu}\right]^{-\frac{\nu+n}{2}},
	\end{align}
	where $\left|\mathbf{C}_k\right|$ is the determinant of $\mathbf{C}_k$.
	
	The shape of the kernel as a function of $x$ is primarily controlled by the correlation structure in $\mathbf{C}_k$, which determines how the kernel varies in different directions of $\VS$. The choice of kernel function mainly determines the rate of decay for large distances $d \gg h$. The Gaussian kernel decays exponentially with $\exp{\left[-(d/h)^2/2\right]}$, while the ST kernel decays as $K_k(d/h) \propto (d/h)^{-(n+\nu)}$. With $\nu=1$, it is known as the Cauchy distribution and has the longest tail among all choices.
	
	The approximation is constructed using a sample $\vec{x}_{L_i}$ of $\pi(x)$. As a result, most points in $\vec{x}_{L_i}$ will be from the high-probability region of $\VS$, leading to the underrepresentation of the tails of $\pi(x)$. If $\pi(x)$ decays quickly away from the most probable region, the Gaussian kernel is appropriate. However, if $\pi(x)$ has long tails or plateaus, the approximation $\tilde{\pi}\condarg{x}{\vec{x}_{L_2}}$ will have significant errors in those regions. To address this, a long-tailed kernel like the Cauchy kernel can be used. Additionally, Cauchy kernels have an advantage in that they lack well-defined mean, variance, and higher moments. This leads to proposal points that can be far from their location, acting as a probe to uncover disconnected regions of high probability in $\pi(x)$ for the MCMC algorithm.
	
	\subsubsection{Weights Choice}
	\label{sec:interpopt}
	
	The kernel density estimation method, first introduced by~\cite{Rosenblatt1956, Parzen1962}, involves the use of kernel functions, which are probability distributions, to estimate the target probability distribution. According to~\cite{Silverman2018}, for a $n$-dimensional function $\pi(x)$, the multivariate density estimation is given by Eq.~\eqref{defapprox} with $w_k = 2 / L$. This method is straightforward and computationally efficient, but its accuracy depends on the representativeness of the sample points $\vec{x}_{L_i}^t$ at time $t$, as it does not use any information about $\pi(x)$ other than these points.
	
	The second approach involves the use of RBI interpolation, which uses the approximation~\eqref{defapprox} to approximate the actual distribution function $\pi(x)$. During the MCMC algorithm, we must compute the posterior at $\vec{x}_{L_i}^t$ to determine the acceptance probability~\eqref{defalpha}. To take advantage of this, we can compute the weights $w_k$ by solving the following system:
	\begin{align}
		\tilde{\pi}(x_i) = \pi(x_i), \quad \text{for} \quad i = 1, 2, ..., L/2.
	\end{align}
	This requires solving the linear system
	\begin{align}
		\label{linsys}	\mathbf{K}\pmb{w} &= \pmb{\pi}, \\ 
		\label{linsyscmp}	\mathbf{K}_{ij} &= \frac{1}{h^n}K_j\mleft[\frac{D_j(x_i, x_j)}{h}\mright], & \pmb{w}_j &= w_j, & \pmb{\pi}_i = \pi(x_i),
	\end{align}
	where the second line defines the components of the matrix $\mathbf{K}$ and vectors $\pmb{w}$ and $\pmb{\pi}$.
	
	The challenge of solving Eq.~\eqref{linsys} is to find a solution that satisfies the constraint $w_i \geq 0$ since the $w_i$ values represent probabilities in~\eqref{defapprox}. This is a Non-Negative Least Squares (NNLS) problem that we addressed using a C algorithm based on the algorithm described in~\cite{Bro1997} and the Lapack Library for linear algebra computations. Note that the weights in regions of low probability have larger errors in the solution of Eq.~\eqref{linsys} as they contribute less to the least squares. To improve the solution, a weighted least squares approach can be used, incorporating weights of $1/\pi(x_i)$. Additionally, interpolation may provide a more precise approximation in higher-probability areas compared to those with lower probability.
	
	\subsubsection{Bandwidth Selection}
	\label{sec:bandwidth}
	
	The final step in calculating the approximation (given by equation~\eqref{defapprox}) involves determining the bandwidth $h$. To do this, we utilize the Rule-of-Thumb (RoT) for density estimation interpolation described in~\cite{Silverman2018}. For Gauss and ST kernels, the RoT is straightforward and simply given by:
	\begin{align*}
		h_\mathrm{RoT}^\mathrm{G} &= \left[\frac{4}{L/2(n+2)}\right]^{\frac{1}{n+4}}, \\
		h_\mathrm{RoT}^\mathrm{ST} &= \left[\frac{16 (\nu - 2)^2 (1 + n + \nu) (3 + n + \nu)}{(2 + n) (n + \nu) (2 + n + \nu) (n + 2 \nu) (2 + n + 2 \nu) L/2}\right]^{\frac{1}{n + 4}}.
	\end{align*}
	This offers a preliminary estimate for the bandwidth. The final bandwidth for both the same and variable kernel approaches is calculated by setting $h = o h_{RoT}$ for the same kernel approach, and $h = o h_{RoT} / p$ for the variable kernel approach, where $p$ is the sample fraction as discussed in \autoref{sec:dist} and $o$ is the over-smoothing parameter. The value of $o$ can be freely chosen by the user or fitted by the algorithm as described below. 
	
	A cross-validation procedure is available with RBF interpolation. The process involves dividing the sample $\vec{x}_{L_i}$ into two subsets: one for the left-hand side of~\eqref{linsys}, containing only kernels with locations on the subset, and the full sample for the right-hand side. In other words, given a split-fraction parameter $f\in(0,1)$ we define the sub-sample $\vec{x}_{fL_i}$ containing the first $m = \lceil f\times L/2 \rceil$ points of $L_i$. The approximation is obtained setting $\pmb{S} = \vec{x}_{fL_i}$ in Eq.~\eqref{defapprox}, where the covariances are still computed using the full sample $\vec{x}_{L_i}$. Now, for each value of $o$, we computed the function 
	\begin{equation}
		C(o) = \left\|\mathbf{K}_f\pmb{w} - \pmb{\pi}_f\right\|^2,
	\end{equation}
	where $\|\|$ represents the L2 norm. The matrix $\mathbf{K}_f$ and vector $\pmb{\pi}_f$ are computed as in Eq.~\eqref{linsyscmp} but with $i=1,2,\dots m$ and $j=1,2,\dots,L/2$. Finally, $C(o)$ is minimized with respect to $o$. The cross-validation method can increase the computational cost of the MCMC algorithm, so it's often more efficient to use a fixed over-smoothing parameter $o$. After some iterations, the algorithm can be stopped to compute the cross-validation estimate for $o$, and then continue the chain using this updated value, rather than performing cross-validation at each step.
	
	\section{Computational tests and Discussion}
	\label{comptests}
	
	In this section, we provide a detailed analysis of the quality of samples generated by the APES algorithm. To evaluate the algorithm's performance, we compare it to the stretch-move algorithm as implemented in \texttt{Emcee}~\citep{Goodman2010, Foreman-Mackey2013}, which we refer to as Stretch. All the necessary components for running the APES algorithm are available in the Numerical Cosmology (NumCosmo) library~\citep{Vitenti2012c}.
	
	To assess the efficiency of the APES algorithm, we use two metrics: the autocorrelation time ($\tau$) and the acceptance ratio of the chains. The autocorrelation time $\tau$ is a measure of the number of iterations required for the next sample point in the chain to become independent~\citep{Goodman2010}. We use the precise definition of $\tau$ for an ensemble sampler given in \cite{Goodman2010} to estimate its value for the APES algorithm. The acceptance ratio of the chains reflects the percentage of proposed moves that are accepted by the algorithm. Finally, to simplify notation, we represent the mean of a quantity given a sample with an overline.
	
	\begin{table}
		\centering
		\caption{NumCosmo approximation options. The Robust column relates to the option to use the robust covariance estimation based on the OGK algorithm. The Interpolation column relates to the option to use RBI interpolation to approximate the desired distribution as discussed in \autoref{sec:interpopt}.}
		\label{tab:appopt}
		\begin{tabular}{lccc}
			\toprule
			Approximation Options & Kernel & Robust & Interpolation \\ 
			\midrule
			\ncopt{KDE} & Same & - & - \\
			\ncopt{VKDE} & Variable & - & - \\
			\ncopt{Robust-KDE} & Same & X & - \\
			\ncopt{Robust-VKDE} & Variable & X & - \\
			\ncopt{Interp-KDE} & Same & - & X \\
			\ncopt{Interp-VKDE} & Variable & - & X \\
			\ncopt{Interp-Robust-KDE} & Same & X & X \\
			\ncopt{Interp-Robust-VKDE} & Variable & X & X \\
			\bottomrule
		\end{tabular}
	\end{table}
	
	NumCosmo provides additional diagnostics for analyzing chain convergence, which we also utilize. We automatically compute the autocorrelation time using NumCosmo and verify the results and diagnostics using GetDist~\citep{Lewis2019}. Sokal~\citep{Sokal1997} provides a detailed discussion of the autocorrelation time and its role in assessing the convergence of the chains and the efficiency of the algorithm.
	
	In addition to examining the autocorrelation time and acceptance ratio, we also compare the analytical mean and variance of the parameters for some distributions with those estimated from the chains. It is worth noting that a high acceptance ratio does not necessarily imply a low autocorrelation and vice versa. A high autocorrelation and acceptance ratio may indicate that the sampling algorithm is not well-adjusted and is proposing new points that are too correlated with the current position. Therefore, we carefully evaluate both parameters to ensure the reliability of the generated samples.
	
	The NumCosmo package offers a variety of kernel functions for density estimation, including Gaussian (Gauss), Student's t with $\nu=3$ (ST3), and Student's t with $\nu=1$ (Cauchy). Each kernel has its own properties, which are discussed in \autoref{sec:K}.
	
	Users can customize their density estimation approach by combining different kernel types with various interpolation and robustness options. For example, the \ncopt{Interp-Robust-VKDE:Cauchy} approach combines variable kernel density estimation (VKDE) with robust covariances and radial basis function (RBF) interpolation using Cauchy kernels. This approach provides the benefits of robust statistics, the flexibility of interpolation, and the heavy-tailed properties of Cauchy kernels. However, it has an increased computational cost.
	
	Alternatively, the \ncopt{KDE:Gauss} approach uses kernel density estimation (KDE) with equal weights and Gaussian kernels, which is a simpler option that is still effective for many applications. Other options, such as \ncopt{VKDE:ST3} that uses the t-distribution kernel with heavier tails than the Gaussian kernel, offer even more flexibility and robustness at a moderate computational cost.
	
	A quick reference guide with labels for each option, along with the kernel, robustness, and interpolation options used, is provided in Tables~\ref{tab:appopt} and \ref{tab:kerntype}. All interpolation options can be combined with every kernel type, allowing users to fine-tune their density estimation approach to meet their specific needs. Bellow we discuss some scenarios where each option in \autoref{tab:appopt} should best perform.
	
	The \ncopt{VKDE} option can be a better fit for distributions with local variations and not centralized around a high probability mode, whereas the \ncopt{KDE} is best suited for centralized and well-behaved distributions. Regarding the ``Interpolation`` option, as discussed in the previous section, it performs better for high probability areas in the parameters space with the downside of being computationally costly. Thus this option can be turned off when exploring areas poorly distributed by the desired probability function. Lastly, the ``Robust`` option is designed for problems that present outliers in the parameter space. This option is also efficient to minimize the effect of higher scales dominating over small scales when computing the Mahalanobis distance in multi-dimensional problems.
	
	Finally, we have introduced a simplified interface for APES that offers compatibility with \href{https://github.com/dfm/emcee}{\texttt{Emcee}} and \href{https://github.com/minaskar/zeus}{\texttt{Zeus}}, allowing users to utilize APES without the need for the complete NumCosmo statistical framework.\footnote{This interface is currently available on the \textsf{master} branch and will be included in the next release.} We have provided an example in the repository demonstrating how to use all three frameworks together.\footnote{\href{https://github.com/NumCosmo/NumCosmo/blob/master/notebooks/apes_tests/rosenbrock_simple.ipynb}{notebooks/apes\_tests/rosenbrock\_simple.ipynb}}
	
	\begin{table}
		\centering
		\caption{NumCosmo kernel types for density estimation.}
		\label{tab:kerntype}
		\begin{tabular}{lcc}
			\toprule
			Kernel Type & Density Function & Properties \\
			\midrule
			Gauss & Eq.~\eqref{kg} & Light-tailed \\
			ST3 & Eq.~\eqref{kt} $\nu=3$ & Heavier tails than Gaussian \\
			Cauchy & Eq.~\eqref{kt} $\nu=1$ & Heaviest tails \\
			\bottomrule
		\end{tabular}
	\end{table}
	\subsection{The 2-dimensional Rosenbrock Distribution}
	
	The first test for the sampler was to generate samples from the 2-dimensional Rosenbrock distribution (see~\cite{Pagani2020} for a discussion on the use of n-dimensional Rosenbrock distributions to test MCMC algorithms). The distribution is given by
	\begin{align}
		\label{eqRB}
		\pi(x_1,x_2) &\propto \exp\left[-100\left(x_2-x_1^2\right)^2+(1-x_1)^2\right], \; x_1,\,x_2 \in \mathbb{R}.
	\end{align}
	where we did not include the normalization. The Rosenbrock distribution is known to be hard to sample from due to its thin tails. Before testing the APES algorithm, we study the interpolation of this distribution using the packages implemented in NumCosmo. Thus, we first generate perfect samples from the Rosenbrock distribution and compute the approximation using different kernels and options.  
	
	\begin{table}
		\caption{Relative difference, acceptance ratio, and over-smooth for each interpolation applied to the same sample of 160 draws from the Rosenbrock distribution.}
		\label{fig:RBfit}
		\centering
		\begin{tabular}{|l|c|c|c|}
			\hline
			Interpolation & $|\tilde{\pi}/\pi-1|<0.2$ & $\overline\alpha$ & $o$ \\ \hline\hline
			\ncopt{Interp-KDE:Cauchy} & 9\% & 64\% &  0.05 \\ \hline
			\ncopt{Interp-KDE:ST3} & 23\% & 65\% &  0.07 \\ \hline
			\ncopt{Interp-KDE:Gauss} & 32\% & 66\% &  0.05 \\ \hline
			\ncopt{Interp-VKDE:Cauchy} & 9\% & 69\% &  0.08 \\ \hline
			\ncopt{Interp-VKDE:ST3} & 33\% & 72\% &  0.13 \\ \hline
			\ncopt{Interp-VKDE:Gauss} & 51\% & 76\% &  0.14 \\ \hline
		\end{tabular}
	\end{table}
	
\begin{figure*}
		\caption{We generated plots of approximated distributions using 160 points distributed according to Eq.~\eqref{eqRB}. For each approximation, we also plotted the $4\sigma$ ellipse in red, which represents the covariance of each kernel. The first line shows plots of the approximated distribution using \ncopt{Interp-KDE} with all supported kernels. In the second line, we used \ncopt{VKDE} with $p=0.05$ (as explained in \autoref{sec:dist}) of the nearest sample points to compute the kernel covariance for each point. All samples were generated with an over-smooth factor $o$ obtained by applying split cross-validation with $60\%$ of the sample ($f=0.6$, as described in \autoref{sec:bandwidth}). Finally, the third line shows a plot of the original Rosenbrock function for comparison. We renormalized all plots so that the maximum probability is equal to one, and we displayed the probability decay until a threshold of $10^{-8}$.	The interpolation with heavier tails tends to overestimate the probability away from the peaks, which is a desirable feature to enable our MCMC algorithm to explore the entire parametric space. Note that the fixed kernel approach is unable to adapt to the local features of the distribution and therefore fails to describe the tails properly. This highlights the advantages of using the VKDE approach to capture the intricate features of the Rosenbrock distribution.}
		\label{fig:RBapprox}
		\centering
		\includegraphics{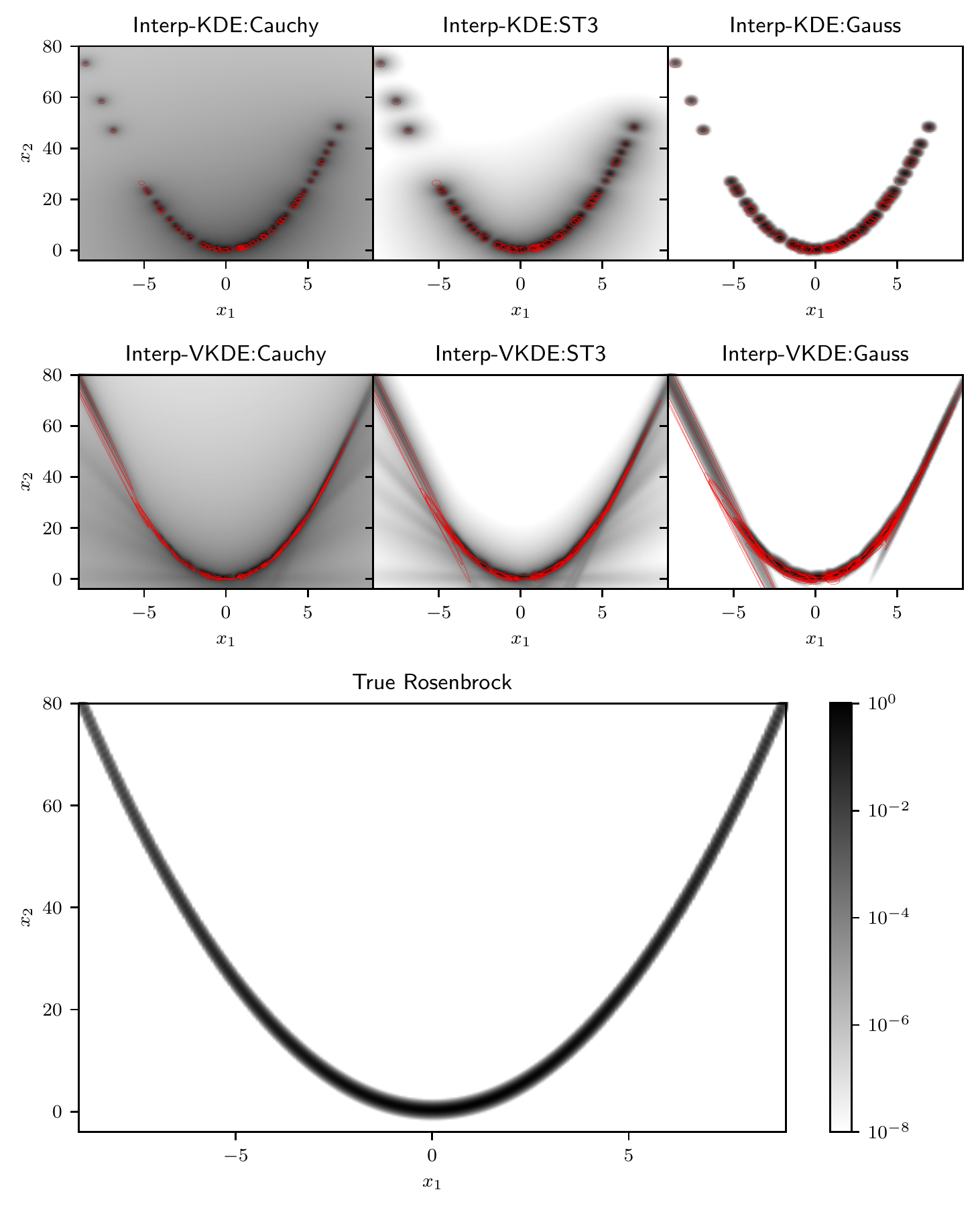}
	\end{figure*}

	In \autoref{fig:RBapprox}, we present different approximations of~\eqref{eqRB}, all of which were generated using the same sample of 160 draws from~\eqref{eqRB}. The VKDE method is found to be superior to the KDE method in accurately describing the target distribution due to its ability to adapt the covariance of the kernels locally. As shown in the first line of the plot, where all ellipses representing the kernel covariances are equal and adapted to the full sample covariance, the position-dependent correlation between $x_1$ and $x_2$ is not accurately captured.
	
	To test the approximation we generated a test sample $\pmb{T}$ from~\eqref{eqRB} with 100,000 points. \autoref{fig:RBfit} shows the percentage of points where the approximation had a relative error smaller than 20\%. Furthermore, we also computed the mean acceptance probability\eqref{acceptprob}. To obtain this probability, we compute 
	\begin{equation}
		\alpha_k\mleft(\pmb{S}, x_{2k}\mright) = \mathrm{MIN}\left[1, \frac{\tilde{\pi}\condarg{x_{2k+1}}{\pmb{S}}\pi\mleft(x_{2k}\mright)}{\tilde{\pi}\condarg{x_{2k}}{\pmb{S}}\pi\mleft(x_{2k+1}\mright)}\right],\quad x_{2k} \in \pmb{T},
	\end{equation}
	obtaining 50,000 values of $\alpha$ and computing its mean $\overline{\alpha}$. The Gaussian kernels yield the smallest errors when fitting~\eqref{eqRB}. Moreover, the mean acceptance probability also follows the same pattern. However, it is important to note that the values in the table are based on a sample of~\eqref{eqRB}, which has higher density in regions of higher probability.  Therefore, we can conclude that Gaussian kernels are better suited to fit these regions of the distribution. 
	
\begin{figure}
		\caption{The upper panel of the graph displays the percentage of test points for which the relative error was smaller than 20\% in bins of $\pi(x)$. It is important to note that all approximations experience a loss of precision as they approach the tails, with lighter-tailed kernels performing better near the peak but suffering from decreased precision in the tails. In contrast, heavier-tailed kernels demonstrate greater stability across the entire range, providing a better overall approximation. To explore the relationship between approximation precision and proposal effectiveness, we have included a plot of the mean acceptance probability in the figure below (see equation~\eqref{acceptprob}). The acceptance probability is shown as a function of the true distribution on the proposed point. Notably, heavier-tailed kernels have the highest acceptance probability in the tails, indicating that these kernels provide a more effective proposal and increase the likelihood of sampling in the tails.}
		\label{fig:RBerralpja}
		\centering
		\includegraphics{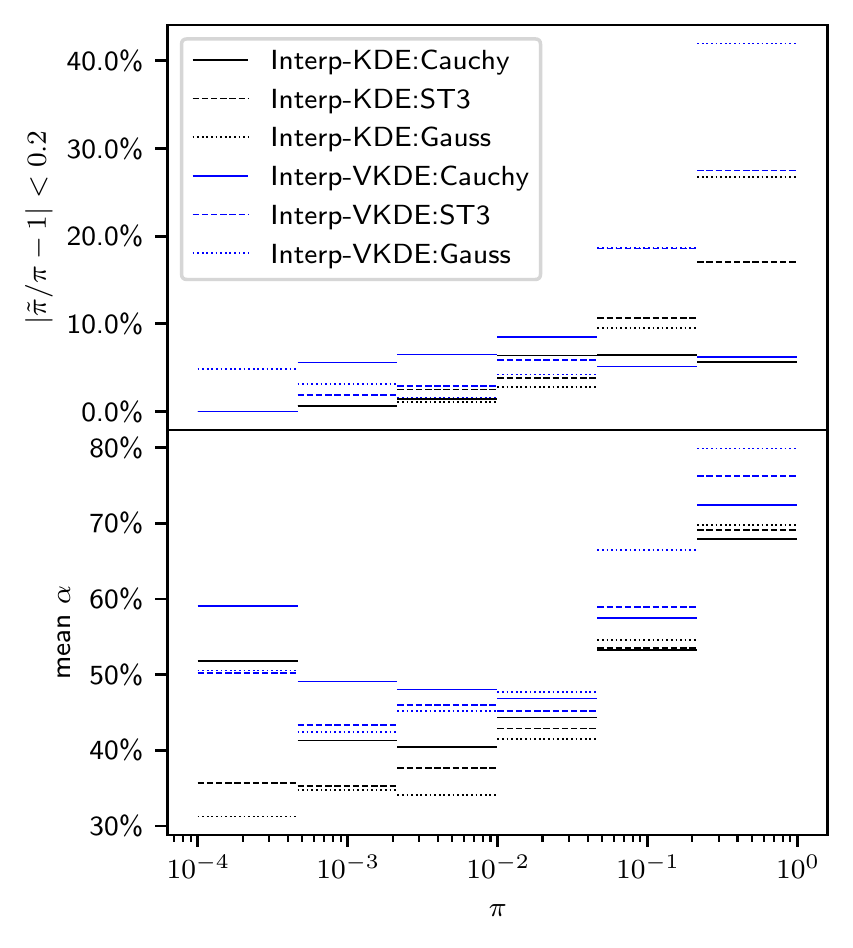}
	\end{figure}

	To gain a better understanding of the errors and the performance of different kernels on the peaks and tails of the distribution, we repeat the same calculation but first we binned the test sample on the values of $\pi(x)$ for $x\in \pmb{T}$ to compute the relative errors and we binned on the target probability $\pi(x_{2k})$ to compute the mean $\alpha$. The results can be seen on \autoref{fig:RBerralpja}. This plot allows us to differentiate the approximation error on peaks and tails, which is crucial for selecting a proposal that can sample the tails effectively.
	
	Our analysis reveals that although Gaussian kernels provide an overall better approximation, they are less accurate on the tails compared to Cauchy kernels. Moreover, the overall performance of the Cauchy kernels is not significantly worse than that of the Gaussian kernels. Therefore, we conclude that Cauchy kernels are better suited to explore the tails of the distribution.
	
	We executed both algorithms using $15625\times320 = 5,000,000$ points, where $15625$ represents the number of iterations and $320$ the number of walkers. In both cases, we discarded the initial 5000 iterations to allow the chains to reach a stationary distribution before collecting samples for analysis. For the APES algorithm, a smaller burn-in period may have been sufficient, but we removed the same number of initial iterations to ensure a fair comparison of the results between the two algorithms. This approach also allowed us to present a clear and consistent analysis of the performance of both algorithms. 
	
\begin{figure}
		\caption{Plot showing the iterations $(x_1^t, x_2^t)$ of the MCMC samples generated from the Rosenbrock distribution using APES and Stretch proposals. The plot includes three chains (out of 320 possible) from each algorithm, where the red points correspond to the APES sample, and the blue points correspond to the Stretch sample. The iterations shown in the plot were obtained after discarding the burn-in period. The visual inspection reveals that the APES algorithm explores the tails more efficiently, with more points distributed away from the mean and exhibiting only small autocorrelation. On the other hand, the Stretch algorithm shows a stronger autocorrelation, with the generated points tending to stay closer to their previous positions. The statistical properties for these two runs are summarized in \autoref{RBstats}.}
		\label{RBparamevol}
		\includegraphics{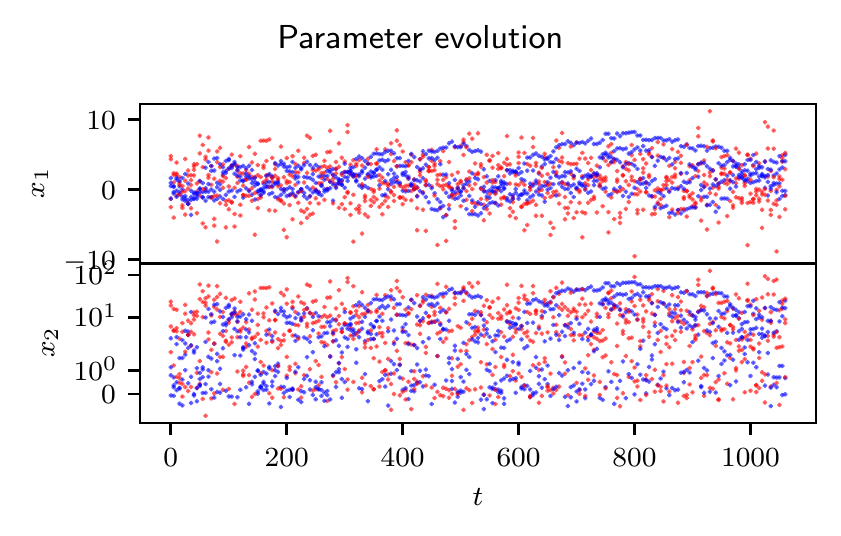}
		\centering
	\end{figure}

	Let us analyze the results in \autoref{RBstats} and Figure~\ref{RBparamevol}. The Rosenbrock sample generated by the APES converged for the true variance value for the Rosenbrock distribution ($\mathrm{Var}\left(-2\ln \pi\right)=4$), while the Stretch sample seems to have not converged yet, as shown in Figure~\ref{RBparamevol}. The autocorrelation time of the APES sample was significantly smaller than the Stretch sample. While the APES provided a time around $\tau = 10$, the Stretch move had an autocorrelation of the order of $\tau = 1400$, and therefore the sample generated by the APES is more independent. One important issue that may be addressed is the acceptance ratio of the APES. Not only APES had a larger acceptance ratio of $47\%$ when compared to the Stretch's acceptance of $23\%$, but the points that are being accepted are more independent than the ones from the stretch sample. The combined effect ends up making APES more efficient than Stretch by a margin of $140$ less autocorrelated.
	
	One important limitation of APES is its computational cost. Building a new approximation of the posterior at each step makes it much more costly than Stretch. In practice, the cost of the Stretch move is zero, while APES requires around 3 minutes to compute all iterations necessary to produce 5,000,000 points when using 320 walkers. Despite its cost, APES has shown to be more efficient than Stretch, producing a sample with lower autocorrelation and better exploring the tails of the distribution. While APES can be improved to decrease its computational cost, for now, we leave that as a topic for future work.
	
	\begin{table}
		\caption{Results of an ensemble sampler run using APES and the Stretch move to generate a 2-dimensional Rosenbrock sample. The APES configuration is \ncopt{Interp-VKDE:Cauchy} with an over-smooth parameter of 0.2 and a local fraction of walkers of 0.05. Both samples were generated using a burn-in of 5000 iterations and a total of 1.6 million points (5000 iterations multiplied by 320 chains). The mean variance and other mean results were computed using all chains. Analytical results were also included for comparison.}
		\label{RBstats}
		\centering
		\begin{tabular}{ |lrrr|}
			\hline
			\multicolumn{4}{|c|}{2-dimensional Rosenbrock Run} \\
			\hline
			& APES & Stretch & Analytical~\eqref{eqRB}\\
			\hline
			Number of walkers   &  $320$   & $320$ & -- \\
			Number of points & $5\times 10^6$     & $5\times 10^6$  & -- \\
			$\tau_{x_1}$ &$6.3$  &  $ 1437.8$ & -- \\
			$\tau_{x_2}$ &$10.7$ &  $ 1421.6$ & --\\
			$\bar{x}_1$  & $1.0$ & $ 0.9 $  & $1$  \\
			$\bar{x}_2$ &$11.0$ & $10.2$  & $11$ \\
			$-2\overline{\ln\pi}$ & $2.0$ & $2.0$  & $2$ \\
			$\mathrm{Var}\left(x_1\right)$  & $10.0 $ & $9.5$ & $10$  \\
			$\mathrm{Var}\left(x_2\right)$  & $236.7 $ & $173.3$ & $2401/10$  \\
			$\mathrm{Cor}\left(x_1, x_2\right)$  & $0.41 $ & $0.35$ & $20/49\approx 0.41$  \\
			$\mathrm{Var}\left(-2\ln \pi\right)$  & $4.0 $ & $3.6$ & $4$  \\
			Acceptance ratio    & $47\%$&  $23\% $ & -- \\
			\hline
		\end{tabular}
	\end{table}
	
	You can find the notebook exploring the approximations of the Rosenbrock distribution,\footnote{\href{https://github.com/NumCosmo/NumCosmo/blob/master/notebooks/stats_dist_tests/stats_dist_rosenbrock.ipynb}{notebooks/stats\_dist\_tests/stats\_dist\_rosenbrock.ipynb}} and the notebook with the MCMC runs and their analysis at NumCosmo github project.\footnote{\href{https://github.com/NumCosmo/NumCosmo/blob/master/notebooks/apes_tests/rosenbrock_mcmc.ipynb}{notebooks/apes\_tests/rosenbrock\_mcmc.ipynb}}
	
	\begin{figure}
		\caption{Corner plot of MCMC runs using APES and Stretch for the Gaussian mixture model defined in Eq.~\eqref{GM2D}. The runs were configured and their burn-ins were determined as described in \autoref{GM2Dstats}. The APES sampler has already converged to the underlying distribution, with the mode on the left having a lower peak due to its larger variance compared to the mode on the right. The bimodal nature of the marginal on $x_1$ poses a challenge in achieving convergence with Stretch, whereas the approximation used in APES can effectively handle multimodal distributions, resulting in a run with low autocorrelation.}
		\label{GM2Dcorner}
		\includegraphics{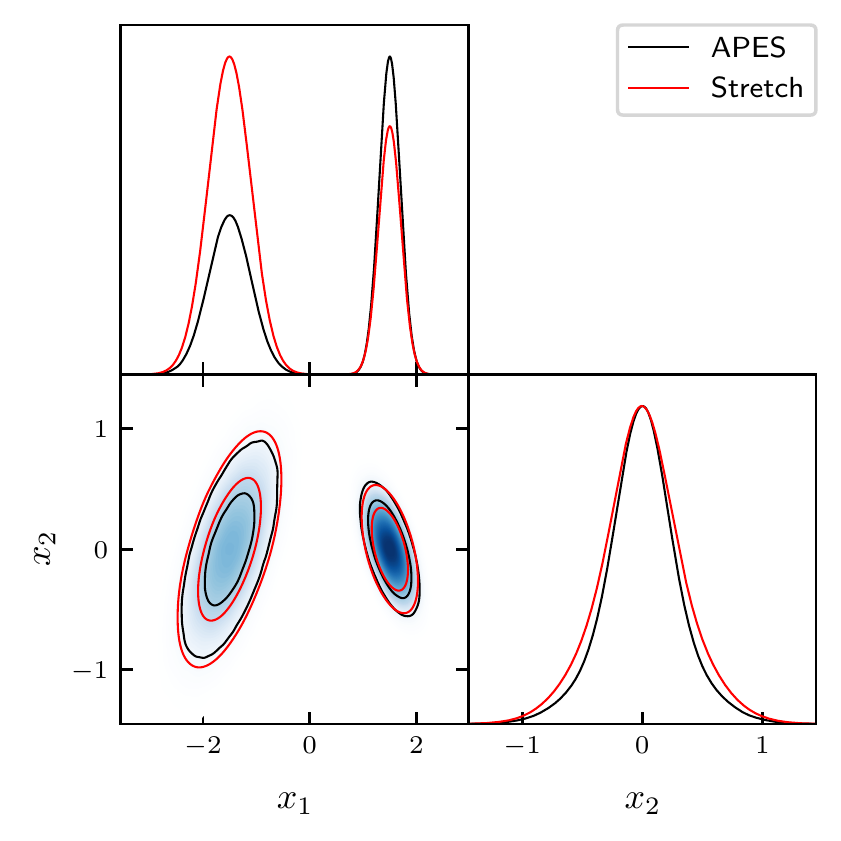}
		\centering
	\end{figure}
	
	\subsection{Gaussian Mixture}
	
	In the previous section, we compared the performance of two algorithms on a unimodal distribution. In this section, we extend our analysis to a multimodal distribution and investigate the ability of APES to explore multiple modes. Specifically, we sample from the Gaussian mixture model given  by
	\begin{equation}\label{GM2D}
		\begin{split}
			\pi(x_1, x_2) = &\frac{1}{2}\mathcal{N}(x_1,x_2, -1.5, 0, 0.4, 0.4, +0.6)\\ 
			+&\frac{1}{2}\mathcal{N}(x_1,x_2, +1.5,0, 0.2, 0.2, -0.6),
		\end{split}
	\end{equation}
	where $\mathcal{N}(x_1,x_2,\mu_1,\mu_2,\sigma_1,\sigma_2,\rho)$ represent a bivariate Gaussian distribution with means $(\mu_1, \mu_2)$, standard deviations $(\sigma_1, \sigma_2)$ and correlation $\rho$.  This distribution consists of two disjoint Gaussian components, each with a different mean and covariance structure. Our goal is to assess how well APES can identify and explore each mode of the distribution.
	
	Again, we executed both the APES and Stretch algorithms using $15625\times320 = 5,000,000$ points and used the same 5000 iterations as burn-in. In \autoref{GM2Dstats}, we present the results of the runs. Note that $x_1$ has a very large autocorrelation only for the Stretch move. This can be understood by inspecting the corner plot in \autoref{GM2Dcorner}, where one can see that the marginal for $x_1$ is strongly bimodal with two prominent peaks. The Stretch move has difficulty with this kind of distribution since it uses a pair of points and makes proposals on the line that connects them. Since these points can be at different parts of the mixture, the proposals end up frequently between the peaks in the valley. The Jupyter notebook used to produce these results can be found at the gihub project page.\footnote{\href{https://github.com/NumCosmo/NumCosmo/blob/master/notebooks/apes_tests/gaussmix2d_mcmc.ipynb}{notebooks/apes\_tests/gaussmix2d\_mcmc.ipynb}}
	
	In this section, we conducted a comparative study of two MCMC algorithms, APES and Stretch, on a Gaussian mixture model with strong bimodal behavior. Our goal was to demonstrate the ability of APES to handle multimodal distributions and compare it with the performance of the Stretch algorithm. The results clearly show that the APES algorithm converges to the underlying distribution much faster than the Stretch algorithm, which can struggle with this kind of distribution due to its use of a pair of points to make proposals. Moreover, the approximation method used in APES has two features that facilitate its handling of multimodal distributions. First, since we use Cauchy kernels, which have heavy tails, APES has the capability of finding the second peak. Second, since we use kernel basis, we can easily handle multiple modes as long as one has enough kernels to approximate both modes. These results demonstrate the effectiveness of APES in handling multimodal distributions and highlight the advantages of the APES algorithm for this type of problem. 
	
	\begin{table}
		\caption{Results of an ensemble sampler run using APES and the Stretch move to generate a sample of~\eqref{GM2D}. The APES configuration is \ncopt{Interp-VKDE:Cauchy} with an over-smooth parameter of 0.2 and a local fraction of walkers of 0.05. Both samples were generated using a burn-in of 5000 iterations and a total of 1.6 million points (5000 iterations multiplied by 320 chains). The mean-variance and other mean results were computed using all chains. Analytical results were also included for comparison. The analytical results with ${}^\star$ were computed using numerical integration.}
		\label{GM2Dstats}
		\centering
		\begin{tabular}{ |lrrr|}
			\hline
			\multicolumn{4}{|c|}{2-dimensional Gaussian Mixture Run} \\
			\hline
			& APES & Stretch & Analytical~\eqref{GM2D}\\
			\hline
			Number of walkers   &  $320$   & $320$ & -- \\
			Number of points & $5\times 10^6$     & $5\times 10^6$  & -- \\
			$\tau_{x_1}$ &$2.2$  &  $ 4017.8$ & -- \\
			$\tau_{x_2}$ &$2.4$ &  $ 48.0$ & --\\
			$\bar{x}_1$  & $0.0$ & $ -0.6 $  & $0$  \\
			$\bar{x}_2$ &$0.0$ & $0.0$  & $0$ \\
			$-2\overline{\ln\pi}$ & $1.57$ & $2.12$  & $1.56^\star$ \\
			$\mathrm{Var}\left(x_1\right)$  & $2.35 $ & $2.02$ & $47/20 \approx 2.35$  \\
			$\mathrm{Var}\left(x_2\right)$  & $0.1$ & $0.12$ & $1/10$  \\
			$\mathrm{Cor}\left(x_1, x_2\right)$  & $0.41 $ & $0.35$ & $20/49\approx 0.41$  \\
			$\mathrm{Var}\left(-2\ln \pi\right)$  & $5.92 $ & $5.63$ & $5.92^\star$  \\
			Acceptance ratio    & $47\%$&  $23\% $ & -- \\
			\hline
		\end{tabular}
	\end{table}
	
	\begin{figure*}
		\caption{Corner plot comparing the performance of MCMC runs using the APES and Stretch methods for the Funnel 10-dimensional model defined in Eq.~\eqref{FD}. The runs were configured and their burn-ins were determined as described in \autoref{FU10stats}. The APES sampler has already converged to the underlying distribution, as indicated by the marginal distribution for $\nu$ which shows the exploration of both the high ($\nu>0$) and low ($\nu<0$) variance regions. In contrast, the marginal distribution obtained using the Stretch move does not accurately reflect the distribution in the low variance region, as it cannot easily explore this region. To account for the large variance difference of the variables $x_n$ conditional to $\nu$, we plotted transformed variables $\sinh^{-1}(x_n)$. Additionally, note that the high variance region is much larger than the low variance region, resulting in a less dense sample in this region. To improve the approximation, more points are needed in the ensemble.}
		\label{FU10corner}
		\includegraphics{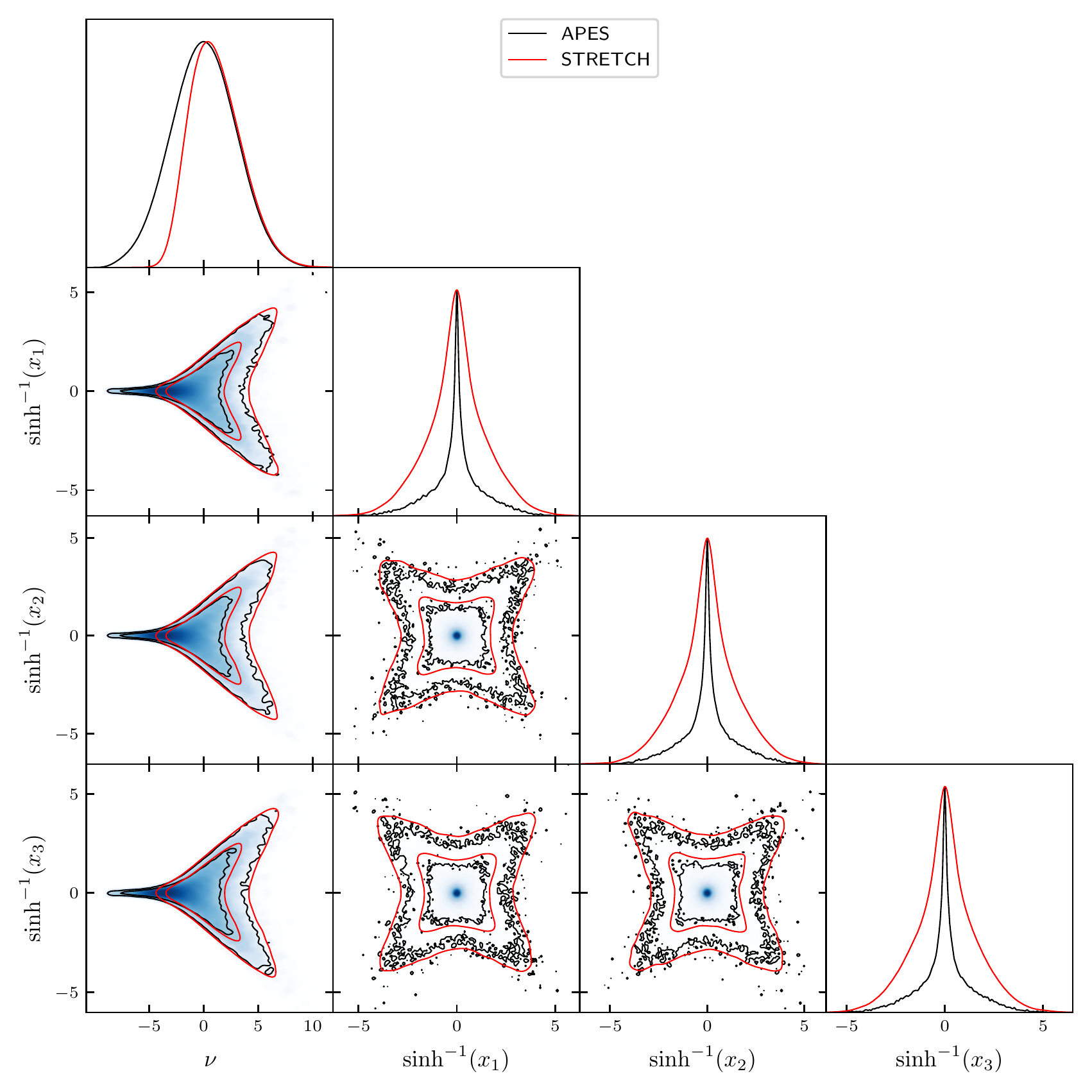}
		\centering
	\end{figure*}
	
	\subsection{The Funnel Distribution}
	
	Our research aims to test the adaptability and effectiveness of our APES method. To achieve this, here we apply our method to the Funnel distribution, which is a well-known benchmark distribution used to validate other sampling methods in the literature~\citep{Neal2003, Thompson2010, Karamanis2020}. The Funnel distribution is particularly challenging due to its complex and multi-dimensional nature. It consists of a Gaussian-distributed variable $\nu$ with a mean of zero and a standard deviation of $\sigma_\nu = 3$, along with $n$ other independent and identically distributed (iid) variables $x_1$ to $x_n$, where their variances $\sigma_{x_i}^2$ are conditional on $\nu$ and determined by $e^\nu$. To obtain samples from the Funnel distribution, we define an $n+1$-dimensional vector $x$ as follows:
	\begin{equation}
		x = (\nu,x_1,,\dots,,x_n)
		.\end{equation}
	The probability density is then
	\begin{align}
		\label{FD}
		\pi(x) &= \frac{\exp \left(-\frac{\chi^2}{2}\right)}{3 (2 \pi )^{\frac{n+1}{2}}}, & \chi^2 &\equiv \frac{\sum _{i=1}^n x_i^2}{e^{\nu }}+\frac{\nu ^2}{9}+\nu n.
	\end{align}
	
	As shown in \cite{Neal2003}, this distribution exhibits a characteristic feature where one region has a low probability density but a high volume of points, while the other region is the opposite. This is due to the large difference in variance for the variables $x_i$ when conditioned to different values of $\nu$. As a result, the Funnel distribution requires a sampler that is adaptive to the local scales of the problem. One such method is slice sampling \citep{Karamanis2020}, which is able to automatically adapt to different scales of a given distribution without requiring any hand-tuning.
	
	Slice sampling has been shown to be a highly effective method for sampling from distributions with complex geometries and varying local scales, such as the Funnel distribution. Our proposed \ncopt{Interp-VKDE:Cauchy} APES method was developed precisely for this type of problem, where traditional MCMC methods can struggle due to the extreme variation in scales across the distribution. To provide a comprehensive evaluation of the performance of our method, we will compare it to the Stretch move.
	
	\begin{table}
		\caption{Data from the ensemble sampler algorithm using the APES and the Stretch move to generate a 10-dimensional Funnel sample. Both samples were generated using a burn-in of $4.2 \times 10^6$ points. The APES sample was generated with the \ncopt{Interp-VKDE:Cauchy} method. Also, we used an over-smooth factor of $0.2$ and local fraction of $0.05$. It was taken the mean value of the results related to the variables $x_1-x_9$ since they have equal moments.} \label{FU10stats}
		\centering
		\begin{tabular}{ |lrrr|}
			\hline
			\multicolumn{4}{|c|}{10-dimensional Funnel Run} \\
			\hline
			& APES & Stretch & Analytical~\eqref{FD}\\
			\hline
			Number of walkers   &  $3000$   & $3000$ & -- \\
			Number of points & $7\times 10^6$     &$7\times 10^6$  & -- \\
			$\tau_{\nu}$ &$33.1$  &  $518.5$ & -- \\
			$\bar{\tau}_{x_1 - x_9}$ & $68.2$ &  $ 172.7 $ & -- \\
			$\bar{\nu}$ & $0.011$ & $1.2$ & $0$ \\
			$\mathrm{mean}(\bar{x}_{n})$ & $0.024$ &  $0.073$ & $0$ \\
			$\overline{\chi^2}$ & $10.1$ & $20.4$  & $10$ \\
			$\mathrm{Var}(\nu)$  & $8.87$ & $ 5.62$  & $9$ \\
			$\mathrm{mean}\left[\mathrm{Var}(x_n)\right]$ & $85.68$ & $171.84$ & $\approx 90.0$ \\
			$\mathrm{Var}\left(\chi^2\right)$  & $740.5$ & $510.9$ & $749$  \\
			Acceptance Ratio    & $10\%$&  $ 36\% $ & -- \\
			\hline
		\end{tabular}
	\end{table}
	
	\autoref{FU10stats} shows the outcomes of the sampling method utilized, including both APES and the Stretch move. It is worth noting that the APES sample was able to depict the underlying distribution described by~\eqref{FD} as one can see in \autoref{FU10corner}. The acceptance ratio underwent considerable changes throughout the sampling process. Initially, as the initial sample did not represent the distribution in~\eqref{FD} well, the kernel interpolation failed to approximate the distribution accurately. During this phase, the acceptance probability in~\eqref{acceptprob} ensured that the sample moved in the direction of higher probability in $\pi(x)$, while the terms involving the approximate distribution were unrelated to $\pi(x)$. This led to a low acceptance ratio of 2\% to 5\%. However, as the sample $\vec{x}$ began resembling a distribution from $\pi(x)$, the acceptance ratio improved to 10\%. It was observed that during the burn-in phase, the interpolation could make the steps less efficient, as it attempted to fit the distribution using points that were not good representatives. Consequently, the covariances were not well adapted to the true distribution. To address this issue, the interpolation was disabled during the burn-in phase, resulting in an improved acceptance ratio. Once the ensemble started to represent the distribution well, the interpolation was re-enabled, leading to better mixing and lower autocorrelation. However, it is important to note that even with the interpolation turned on, the chains eventually converge to a good distribution, albeit with longer chains.
	
	In~\cite{Karamanis2020}, the mean autocorrelation time was found to be $\tau = 129$ for a $25$-dimensional Funnel distribution using ensemble slice sampling. In our experiment with the same $25$-dimensional Funnel distribution, we employed APES with 5000 walkers, resulting in an acceptance ratio of $\approx 1\%-2\%$ and an autocorrelation of $\approx 150$. One challenge of increasing the ensemble size is that it takes longer to move past the burn-in phase, as more points need to be moved to the correct distribution. One possible strategy to address the challenge of a longer burn-in phase with larger ensemble sizes is to employ fewer walkers in the initial sampling phase until a reasonable approximation of the posterior distribution is achieved. This approximation can then be used to generate an initial distribution for further analysis with a larger ensemble. We followed the strategy of increasing the number of walkers from 5000 to 8000 in the analysis of the 25-dimensional Funnel distribution, resulting in a mean autocorrelation of approximately 80 after the burn-in is removed. The Funnel analysis can also be found on the NumCosmo github.\footnote{\href{https://github.com/NumCosmo/NumCosmo/blob/master/notebooks/apes_tests/funnel_mcmc.ipynb}{notebooks/apes\_tests/funnel\_mcmc.ipynb}}
	
	\begin{figure*}
		\caption{The figure below depicts two panels showcasing the performance of the APES algorithm using the \ncopt{Interp-VKDE:Gauss} approach. In the upper panel, we present the acceptance ratio, representing the percentage of proposed points accepted per step, for various choices of the Walkers per Dimension ($w_d$) parameter. The algorithm is executed across a range of dimensions from 2 to 50, in each case, we perform a total of 1000 iterations. On the other hand, the lower panel illustrates the largest autocorrelation time ($\tau$) observed among all parameters. Remarkably, for low dimensions, even a modest $w_d$ value of 100 yields a highly favorable approximation. This is evident from the near 100\% acceptance ratio and nearly negligible autocorrelation (i.e., close to unity). However, as the dimensionality increases, the approximation quality gradually deteriorates. Consequently, the acceptance ratio decreases, while the autocorrelation becomes more pronounced. Despite this behavior, APES is robust up to $n=35$, for higher dimensions the acceptance ratio drops below 5\%, and $\tau$ increases over 50 when $w_d$ is set to 600. Nevertheless, even for $n=35$, the algorithm still generates a desirable posterior sample with an autocorrelation level comparable to other MCMC algorithms. It is worth noting that, as the dimensionality increases, the autocorrelation time approaches the total number of iterations, which is set to 1000. This convergence of autocorrelation and the limited number of iterations make estimating $\tau$ challenging, resulting in a high variance in its estimate. Therefore, it is reasonable to assume that for these cases, additional iterations would be required to accurately estimate $\tau$.}
		\label{GCevol}
		\includegraphics{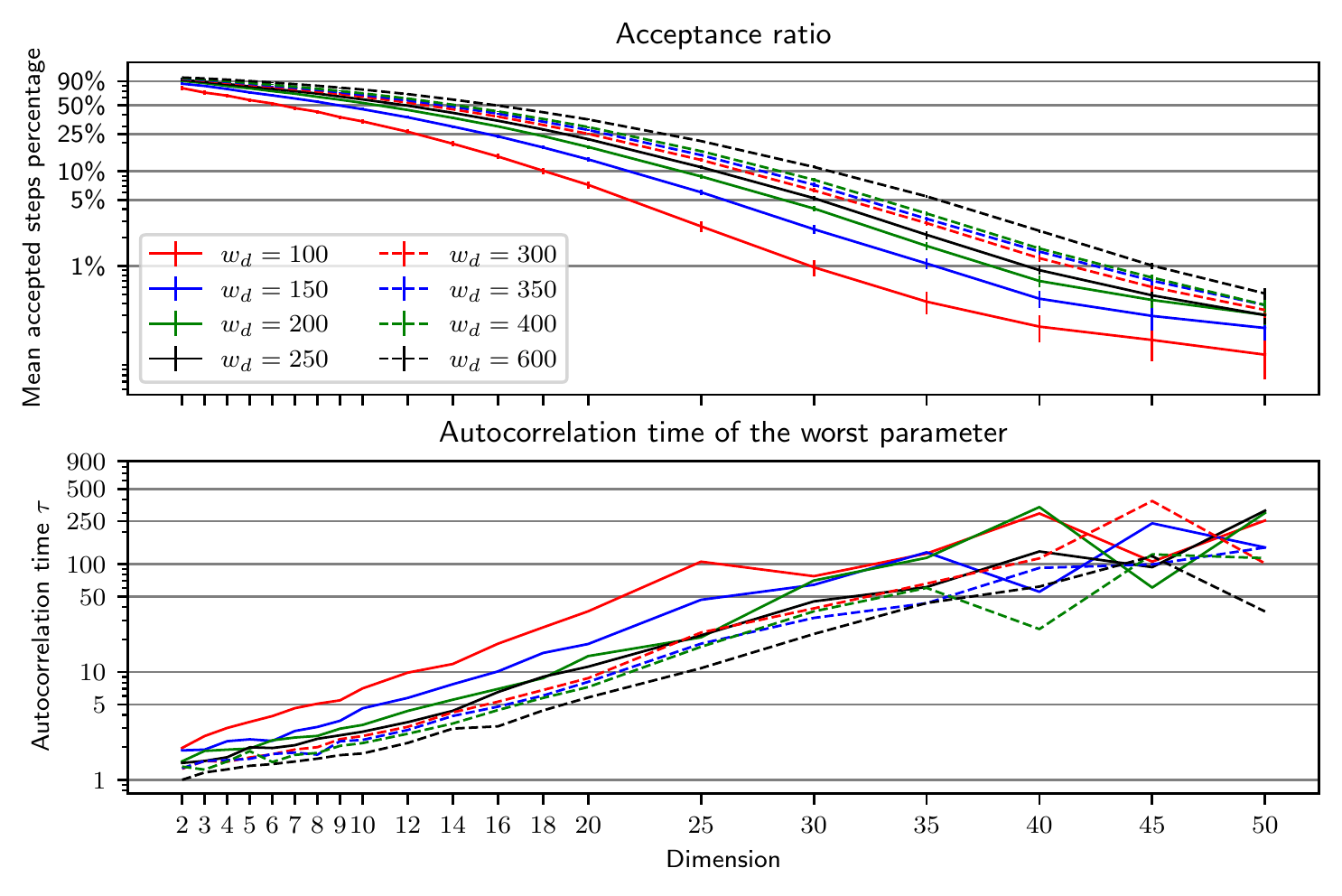}
		\centering
	\end{figure*}
	
	\subsection{Truncated Multivariate Gaussian Distribution}
	\label{TMVN}
	
	To investigate the scaling behavior of the APES algorithm as dimensions increase, we evaluate its performance on a truncated multivariate Gaussian distribution. Specifically, the distribution we consider is defined as follows:
	\begin{equation}
		\label{eqTMVN}
		\pi(x) \propto \exp\left[-\frac{x^\intercal C^{-1} x}{2}\right], \; x \in \mathbb{R}^n, \; x_i \in \left\{\begin{array}{ll}
			(0, 10) & \forall\;i\;\text{odd} \\
			(-10, 10) &  \forall\;i\;\text{even}
		\end{array}\right.,
	\end{equation}
	where $C$ represents the covariance matrix. Our objective is to apply the APES algorithm to this distribution while varying the dimensionality from 2 to 50. To facilitate a comprehensive analysis of how the algorithm scales with dimension, we generate a $50\times50$ covariance matrix with a unit diagonal and random correlations between the variables, denoted as $C_{50}$, once. For each dimension $n$, we extract the corresponding $n \times n$ top-left block from $C_{50}$ and employ it as the covariance matrix for our experiments.
	
	\begin{figure*}
		\caption{The figure below presents a corner plot comparing the benchmark, represented by the independent and identically distributed (iid) distribution of the posterior computed using the code \textsf{truncated-mvn-sampler} (here labeled as TMVN) described in \autoref{TMVN}, with the output obtained from APES. The analysis focuses on the challenging scenario of $n=50$ and $w_d=600$, where the APES sampling has not yet fully converged. For the sake of clarity and to focus on the most informative aspects, we have opted to display only the first eight dimensions in the plot. Including the remaining dimensions would not provide any additional relevant information or insights. To provide a meaningful comparison, we consider the final ensemble generated by APES, comprising $30,000$ points, and compare it to the benchmark distribution. It is worth noting that this particular characteristic of the confidence regions, where only the tail of the multivariate Gaussian distribution is captured, becomes more prominent when combined with a very large dimensional space. In the case of the plane $(x_3, x_5)$, the constraints and correlations result in a confidence region that encompasses only a wedge-like portion of an ellipse, and the corresponding one-dimensional marginal distributions predominantly exhibit the tails. This highlights the challenge of exploring and accurately representing complex regions in high-dimensional spaces using APES.}
		\label{GCcorner}
		\includegraphics{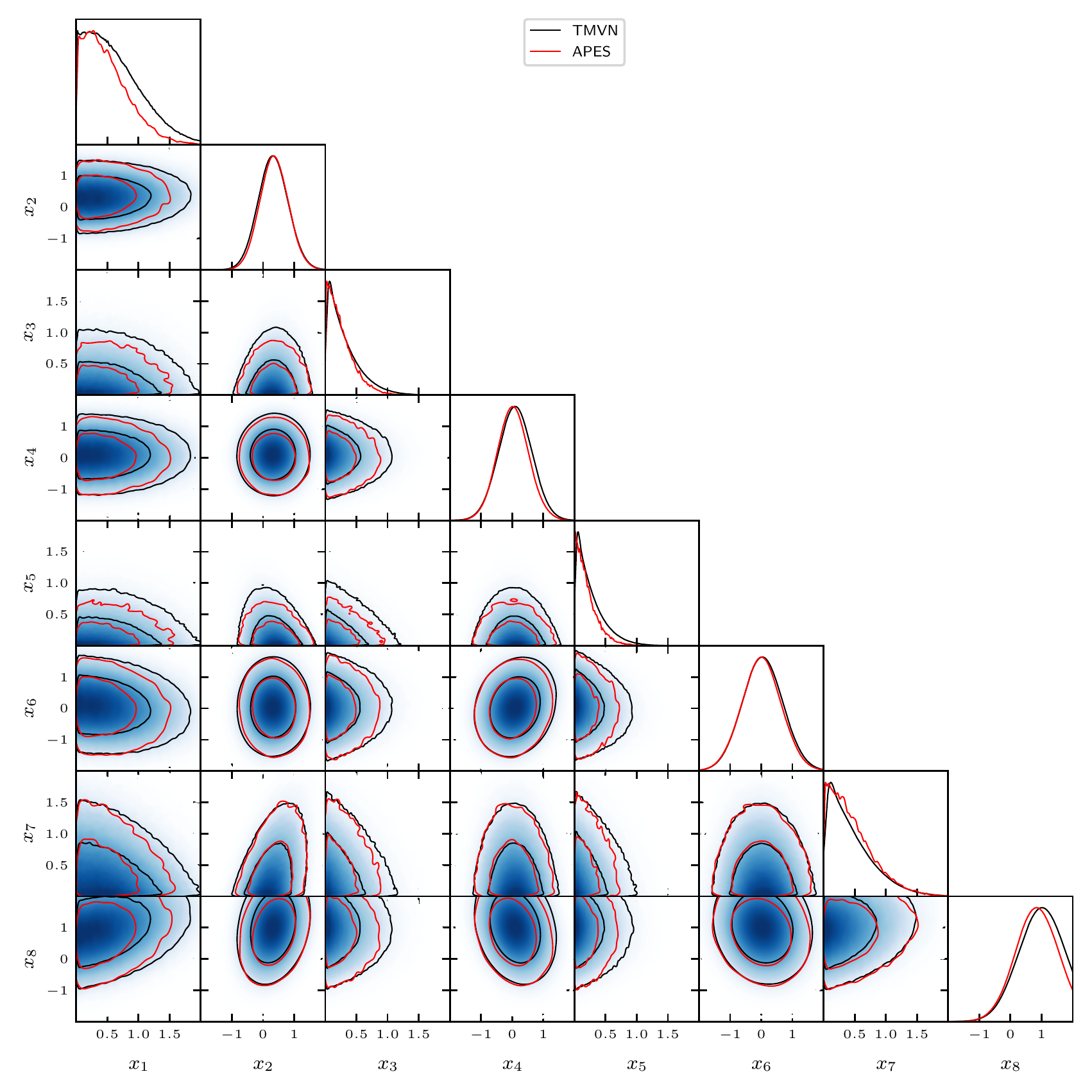}
		\centering
	\end{figure*}
	
	Considering the truncated multivariate Gaussian distribution, it exhibits a combination of thin tails and sharp cuts. This particular distribution is characterized by a rapid decrease in probability density as we move away from the mean, resulting in thin tails. Furthermore, the distribution is truncated, meaning it is limited within specific ranges for each dimension, resulting in sharp cuts at the boundaries. To accommodate these characteristics, we employ the Gaussian kernel within the APES algorithm. In addition, we utilize the VKDE technique in conjunction with interpolation, specifically adopting the \ncopt{Interp-VKDE:Gauss} approach, to effectively capture the distribution's behavior. For all our experiments, we maintain a fixed bandwidth of 1.1 for the Gaussian kernel used in the APES algorithm. Additionally, we set the local fraction to $p=0.05$ for the \ncopt{VKDE} option. For our initial sample, we compute the initial ensemble using an uncorrelated multivariate Gaussian distribution. Each component has a mean of $x_i = 0.5$, and the variance is given by $\sigma_i = 0.1$. This choice allows us to initialize the sampling process within a favorable region of the posterior, although the initial step may not be representative of the target distribution. By starting in a region of reasonable posterior probability, we avoid long burn-in phases. Using this strategy and maintaining a fixed configuration for all other aspects of the algorithm, we aim to investigate APES as a self-contained, black-box approach, enabling us to comprehensively explore its scaling behavior with increasing dimensions.
	
	As the dimensionality of the problem increases, the volume of the probability space expands correspondingly. This growth is typically exponential in terms of the dimension $n$. Consequently, it is expected that larger dimensional problems would necessitate a greater number of walkers (kernels) to adequately approximate the distribution. To address this, we introduce a variable called "walkers per dimension" denoted as $w_d$. This variable quantifies the necessary increase in the total number of walkers as the dimensionality increases. For instance, if we set $w_d = 100$, then when tackling a problem with $n = 5$ dimensions, we would employ $100 \times 5 = 500$ walkers. It is important to note that while the number of walkers scales linearly with $n$ for a fixed $w_d$, the volume of the probability space expands exponentially. Consequently, we anticipate the need to increase the value of $w_d$ as the dimensionality $n$ grows in order to maintain sufficient coverage of the target distribution. 
	
	To assess the accuracy of our results, we utilize a specialized sampling method specifically designed for truncated multivariate Gaussian distributions, as described by~\cite{Botev2016}. This method, implemented in the GitHub repository, \footnote{\href{https://github.com/brunzema/truncated-mvn-sampler}{truncated-mvn-sampler}} provides a reliable framework for sampling from such distributions. By employing this established approach, we can validate the performance and effectiveness of our APES algorithm in approximating the target distribution accurately.

	\begin{table*}
		\caption{Results from the cosmological analysis are presented. The APES data were generated using the \ncopt{Interp-VKDE:Cauchy} option with a local fraction of $p = 0.05$ and an over-smooth factor of 0.2. We generate two samples using 2000 walkers using both APES and Stretch. Additionally, we included a run using Stretch and just 64 walkers. Although the APES chains had converged since the 200th iteration, the larger autocorrelation resulting from the Stretch move necessitated increasing the number of steps to obtain a better approximation of the posterior using this sampler. In this study, we intentionally included the slow-convergence phase of the Stretch move to show its influence and bias on the parameters. Increasing the burn-in to 500 iterations removes the bias and reduces the autocorrelation to about 40 for all parameters. All tests converged to the same means and variances: $\overline{H}_0 = 72.3 \pm 0.9$, $\overline{\Omega}_{c0} = 0.27 \pm 0.04$, $\overline{\Omega}_{k0}= -0.21 \pm 0.05$, $\overline{w} = -0.9 \pm 0.05$, $\overline{m}_{\nu0} = 1.4 \pm 2.0$, and $\overline{\mathcal{M}}_1 = -19.3 \pm 0.03$.} \label{CA}
		\centering
		\begin{tabular}{ |lrrrrr|  }
			\hline
			\multicolumn{6}{|c|}{Pantheon BAO sample} \\
			\hline
			& APES & APES & Stretch & Stretch & Stretch \\
			\hline
			Number of walkers   &  $600$ & $2000$  & $64$ & $600$ & $2000$ \\
			Total number of points & $2 \times 10^{6}$  &$ 2 \times10^{6}$  & $2 \times 10^{6}$  & $2 \times 10^{6}$  & $2 \times 10^{6}$ \\
			Burn-in: steps & 68 & 130 & 64 & 544 & 670 \\
			Burn-in: points & $\approx 4.1\times10^4$ & $\approx 2.6 \times 10^5$ & $\approx 6.0 \times 10^4$ & $\approx 3.3 \times 10^5$ & $\approx 1.3 \times 10^6$ \\
			Effective sample size & $\approx 2.9\times 10^5$ & $\approx 4.7 \times 10^5$ & $\approx 9.9 \times 10^3$ &  $\approx 1.9\times 10^4$ & $\approx 2.6 \times 10^3$\\
			$\tau_{H_0}$ &$3.9$  &  $2.6$ & $103$ & $48$ & $23$ \\
			$\tau_{\Omega_{c0}}$ &$6.4$ &  $3.7$ & $196$ & $71$ & $21$\\
			$\tau_{\Omega_{k0}}$  & $4.9$ & $3.2$  & $80$ & $83$ & $25$\\
			$\tau_{w}$ &$5.2$ & $2.5$  & $87$  & $79$ & $24$\\
			$\tau_{m_{\nu0}}$ &$6.6$ & $3.7$ & $196$ & $86$ & $23$  \\
			$\tau_{\mathcal{M}_1}$  &$3.9$ & $2.6$ & $102$  & $57$ & $24$\\
			Acceptance ratio  & $40 \%$&  $49\% $ & $43\%$  & $43\%$ & $44\%$ \\
			\hline
		\end{tabular}
	\end{table*}
	
	In \autoref{GCevol}, we present the evolution of APES efficiency as the dimensionality increases, considering different values of $w_d$. In all tests, we conduct 1000 steps, resulting in a total number of points in a sample given by $1000 \times n \times w_d$. For instance, in our simplest test with $n=2$ and $w_d=100$, we obtain $1000 \times 2 \times 100 = 200,000$ points. On the other hand, in the most complex scenario with $n=50$ and $w_d=600$, we obtain $1000 \times 50 \times 600 = 30,000,000$ points.
	
	Considering computational costs, as the number of walkers (kernels) increases, the expense of constructing the approximation also escalates. To address this, we implemented a parallel version of the VKDE algorithm to compute the covariances $\mathbf{C}_k$, the interpolation matrix~\eqref{linsyscmp}, and solve the system~\eqref{linsyscmp} concurrently. All tests were conducted on a machine equipped with two 10-core Intel(R) Xeon(R) CPU E5-2640 v4 2.40GHz processors. 
	
	In the worst-case scenario, with $n=6$ and $w_d=600$, it takes approximately 4 minutes using 20 cores to compute the approximation for each block. Consequently, it takes around 8 to 9 minutes to complete each iteration. However, considering the large number of walkers ($30,000$) in this case, the computational time per walker is reasonably small, at approximately 18 milliseconds per posterior computation. The computation time decreases substantially for lower dimensions. For instance, when considering $n=35$ and $w_d = 600$, the iteration time decreases to approximately 3-4 minutes. Similarly, for $n=20$ and $w_d=600$, the iteration time reduces to around 1 minute. Hence, the computational time concern primarily arises for higher dimensional problems.
	
	To provide some perspective, in cosmology, numerous studies involve solving the Boltzmann equations coupled with first-order cosmological perturbations. Typically, this step takes around 4 seconds to complete in a single core. Therefore, in such cosmological scenarios, the computational time required for building the APES approximation would have a small impact on the overall computation time.
	
	However, it is important to note that in cases where the computational cost of evaluating the posterior is relatively low (as in the current example), the cost of APES becomes the dominant factor affecting the total computational time. While the posterior evaluation itself may be efficient, the construction and updating of the APES approximation become more time-consuming due to the increased number of walkers and dimensions. Therefore, in situations where the cost of the posterior evaluation is minimal, careful consideration should be given to optimizing the efficiency of the APES algorithm to mitigate the impact on computational time.
	
	To assess the accuracy of the samples, we generate a corner plot in \autoref{GCcorner}, which includes the benchmark represented by the independent and identically distributed (iid) distribution of the posterior computed using the method developed by~\cite{Botev2016}, as well as the output obtained from APES. In this analysis, we focus on the more complex scenario of $n=50$ and $w_d=600$. Since the APES sampling in this scenario has not completely converged, we only compare the final ensemble, which contains $30,000$ points, with the benchmark.
	
	In this example, it is worth noting that while the APES sample may not fully explore the tails of the distribution, it provides a reasonable approximation of the posterior for most parameters. However, due to the constraints imposed on the odd dimensions and the correlations between parameters, the distribution is truncated and includes only the tail of the otherwise unconstrained posterior distribution. This truncation effect becomes evident when examining the bidimensional confidence regions between odd parameters. These more intricate regions are less explored by APES and would require additional steps to fully converge and accurately capture their characteristics.
	
	\begin{figure*}
		\caption{The figure illustrates the parameter evolution as a function of the iteration $t$, focusing on the burn-in phase and the initial mixing for MCMC runs using the APES and Stretch methods in a cosmological analysis. We specifically examine the most efficient configuration of 600 walkers for both methods. To maintain clarity in the visualization, we included only 20 chains in the plot. The parameter evolution shown pertains to two specific parameters: $H_0$, which exhibits smaller autocorrelation, and $m_{\nu0}$, which exhibits higher autocorrelation. It is worth noting that the burn-in for the Stretch method occurs around $t=544$, as indicated in \autoref{CA}. This can be observed in the figure as it takes this duration for the Stretch method to begin exploring the region with $m_{\nu0} > 0.1$. }
		\label{CAFigure}
		\includegraphics{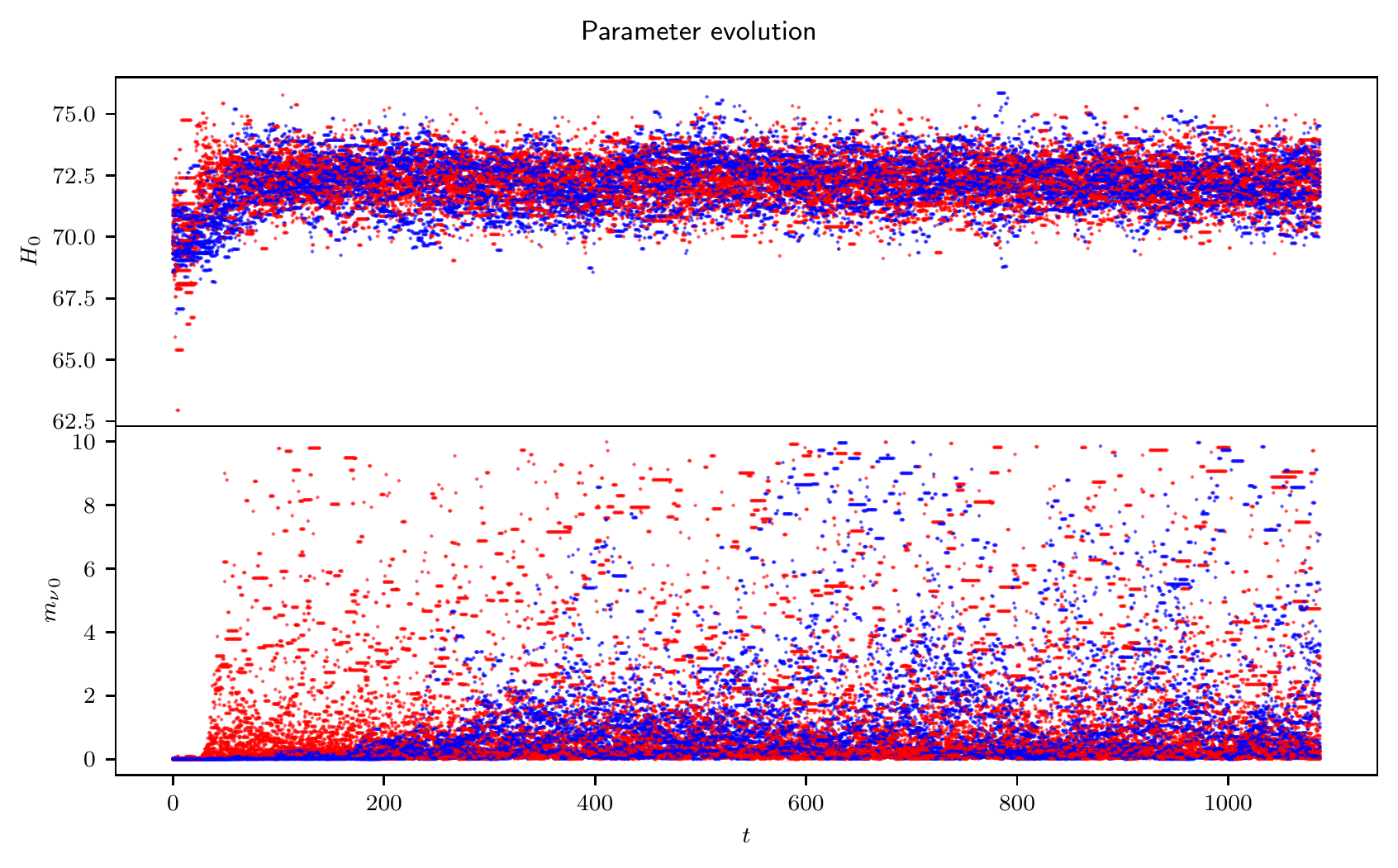}
		\centering
	\end{figure*}
	
	\subsection{Cosmological Model}
	
	This section delves into an examination of the parametric space within the standard cosmological model, which incorporates a single massive neutrino. Our analysis involves fitting several parameters, namely the Hubble parameter $H_0$, the cold dark matter density parameter $\Omega_{c0}$, the curvature parameter $\Omega_{k0}$, the dark energy equation of state $w$, the supernovae type Ia absolute magnitude $\mathcal{M}_1$, and a massive neutrino $m_{\nu0}$ (in electron-volt eV), as described in detail in~\cite{Doux2018}. The six-dimensional parametric space is denoted as $x = (H_0,\,\Omega_{c0},\,\Omega_{k0},\,w,\,m_{\nu0},\,\mathcal{M}_1)$. We adopt simple flat priors for all parameters, which have minimal impact on the results since the fitting region is sufficiently far from the borders, except for cases where $m_{\nu0}>0$ and $\Omega_{k0}>-0.3$.
	The likelihood function used in our analysis comprises the Pantheon+ likelihood~\citep{Brout2022, Scolnic2022}, the Baryon Acoustic Oscillation (BAO) derived cosmological distances obtained from Data Release 16 (DR16) of the SDSS-III~\citep{Alam2016, Alam2021}, which also incorporates all non-overlapping DR12 distances, and the cosmological chronometers collated in~\cite{GomezValent2019}. Our deliberate selection of a poorly constrained parametric space aims to study the behavior of the sampler when dealing with intricate and unconstrained parametric spaces in a real-world scenario.
	
	In NumCosmo APES, we have the option to determine the best-fit prior before running the MCMC, which has been observed to significantly accelerate chain convergence. However, for the purpose of studying the burn-in phases of APES in this analysis, we chose not to use this option. Instead, we initiated all runs with a small cloud of points sampled from independent Gaussian distributions with the following parameters:
	\begin{align*}
		H_0 &= 70 \pm 1, & \Omega_{c0} &= 0.25 \pm 0.01, & \Omega_{k0} &= 0.0 \pm 0.01, \\
		w &= -1 \pm 0.01, & m_{\nu0} &= 10^{-5} \pm 0.01, & \mathcal{M}_1 &= -19.25 \pm 1.
	\end{align*}
	In our analysis, we focus on two important considerations. Firstly, using a larger number of walkers tends to prolong the burn-in phase as more points need to be guided to the appropriate region of the parametric space. Secondly, increasing the number of walkers not only improves the quality of approximations but also reduces the autocorrelation time. It is worth noting that this observation holds true for the Stretch move as well. Although the Stretch move does not require as many walkers as APES, employing only a few walkers in the Stretch move may restrict the steps at each iteration to a limited portion of the parametric space, resulting in higher autocorrelations.
	
	We conducted experiments using five different configurations: APES with 600 walkers, APES with 2000 walkers, and Stretch with 64, 600, and 2000 walkers. In each case, we ran a total of approximately $2 \times 10^6$ points. The posterior parameters and covariance were computed by removing the burn-in phase, which was determined as the iteration that maximizes the effective sample size. The effective sample size is defined as the ratio of the remaining number of points to the autocorrelation time.
	
	As depicted in \autoref{CA}, the burn-in phase exhibits an increasing trend with the number of walkers, which aligns with our expectations. Interestingly, both APES and Stretch with 600 walkers demonstrated larger effective sample sizes, indicating improved convergence due to reduced autocorrelation. This highlights the benefits of employing a larger number of walkers in these configurations.
	
	It is important to note that despite differences in the burn-in phase, all configurations ultimately converged to the same mean and covariance for the parameters of interest. Notably, Stretch with 64 walkers exhibited a significantly shorter burn-in phase. This suggests a potential strategy where one can initially utilize a smaller number of walkers to quickly pass through the burn-in phase and then restart the analysis using the current sample with a larger number of walkers. While we are currently investigating this possibility and exploring the mixing of APES with other methods, such as the Stretch move, we defer these investigations to future work.
	
	In the most efficient configurations (both utilizing 600 walkers), we observe a significant reduction in autocorrelation in the APES sample compared to Stretch. Specifically, the APES sample exhibits approximately 20 times less autocorrelation than Stretch. This distinction becomes visually apparent when examining \autoref{CAFigure}, which illustrates the evolution of 20 different chains.
	
	An interesting observation from the plot is that both methods exhibit similar behavior when the chains are far from the mean. In these regions, the chains tend to remain stuck at a particular value for an extended number of iterations. Additionally, in the case of Stretch, the points frequently cluster together, further increasing the autocorrelation.
	
	Another noteworthy observation pertains to the high autocorrelation observed when utilizing Stretch with 2000 walkers, as depicted in \autoref{CA}. In such cases, the burn-in phase constitutes a substantial portion of the overall iterations, exceeding half of the total. When confronted with this scenario, it becomes crucial to exercise caution in the analysis process.
	
	In these situations, conducting a shorter analysis can result in a noisy estimate of the autocorrelation, as discussed in \autoref{TMVN}, potentially leading to a false sense of convergence. It is important to note that this issue is not exclusive to a particular sampler but applies universally. Whenever the autocorrelation is excessively high, it becomes imperative to approach the results with care.
	
	This is precisely why APES proves to be highly advantageous in numerous scenarios. Its typically low autocorrelation offers a means to mitigate this problem effectively. By minimizing the autocorrelation, APES enables more reliable and robust analyses, reducing the risk of false convergence and facilitating more accurate estimation of the target distribution.
	
	\section{Conclusion}
	\label{conclusion}
	
	In summary, the radial basis interpolation technique employed in the APES algorithm has shown great efficiency in generating approximate posterior distributions and generating sample points from them, with smaller autocorrelation times and good acceptance ratios compared to other MCMC algorithms. Our evaluation of the APES and Stretch sampler using the same posterior distributions demonstrated that the APES algorithm outperformed the Stretch move sampler, particularly in adaptive sampling problems such as the Funnel and Rosenbrock distributions.
	
	However, it is important to note that in the large autocorrelation cases, the chains tend to falsely appear converged early in the analysis. This is because the slow convergence can result in false positives when diagnosing the results, as shown in our experiments. The tails of the distribution are not well-represented by the chains when the slow phase is included. In this phase, the chains are still exploring relevant regions of the parametric space, as evidenced by the large autocorrelation. Moreover, the slow convergence phase can bias the mean and report a lower variance due to the over-representation of the distribution center.
	
	We have also tested the algorithm for a cosmology likelihood, and our results showed that the APES algorithm had a much smaller autocorrelation time when compared to the Stretch move sampler. For wider samples, the APES algorithm is expected to have a smaller autocorrelation, which was confirmed by our experiments. In this test, we also analyzed the efficiency of the algorithm regarding the number of walkers and stated that a large number of walkers prolongs the burn-in phase. However, a similar analysis was made in the truncated multivariate Gaussian distribution test, where we rescaled the number of walkers with the problem's dimension. In this case, the efficiency of the algorithm improved with the number of walkers when avoiding longer burn-in phases by starting at high probabilty areas. We are currently exploring other possibilities to avoid extended burn-in phases.
	
	In the Funnel distribution test, the APES algorithm demonstrated its effectiveness in solving problems that necessitate an adaptive sampler. While the APES algorithm exhibited satisfactory overall performance, there is scope for improvement by integrating a combination of different approximation methods at different stages of the sampling process. Furthermore, in certain situations, the interpolation technique may not be optimal for the burn-in phase or the over-smoothing parameter may require more precise adjustment. Therefore, users should adapt the sampler to suit the specific problem and assess which approach produces the best outcomes. Through this approach, the APES algorithm can offer even more efficient and accurate results for a wide range of applications.
	
	Additionally, we conducted a study to assess the performance of APES as the dimensionality increases in the case of a truncated multivariate Gaussian distribution. Our findings indicate that APES performs well up to $n=35$. However, beyond this point, achieving convergence requires a significantly larger number of iterations. We are currently engaged in ongoing research to enhance APES specifically for high-dimensional scenarios.
	
	One potential approach we are exploring involves exploring the posterior by moving through one subspace at a time, akin to a Gibbs sampler. This strategy aims to mitigate the computational cost associated with constructing approximations in high dimensions. While promising, it is important to note that this approach is still a work in progress and requires further development and refinement. 
	
	We thoroughly tested the robust method in various scenarios and found that while it did not show a significant improvement in the sampling process as a whole, it did come at a significant cost. Despite its potential benefits, the computational expense associated with the robust method makes it impractical for large-scale applications. Therefore, in practical settings where computational efficiency is paramount, our findings suggest that the standard approach may still be preferable. Nonetheless, further research into more computationally efficient robust methods could yield valuable insights for future applications. For instance, during the burn-in phase of the sampling process, robust methods can better approximate the covariance even in the presence of many outliers. This capability can be particularly important in certain applications where accurate estimation of the covariance is critical for the downstream analysis. Therefore, it may be worth considering the use of robust methods during the burn-in phase or in situations where outliers are expected to be prevalent. Nonetheless, it's important to weigh the potential benefits against the added computational cost when making this decision.
	
	The versatility of the APES algorithm enables it to integrate with any available approximation technique, making it a powerful tool for various applications. Its adaptable framework and efficient sampling of the target distribution has been demonstrated to yield faster convergence rates and less autocorrelated results compared to other widely-used MCMC algorithms. Although the integration of additional kernels, cross-validation, and fitting techniques could enhance the framework's capabilities in the future, the current framework is already highly effective and appropriate for solving a wide range of problems.
	
	\section*{Acknowledgements}
	
	SDPV acknowledges the support of CNPq of Brazil under grant PQ-II 316734/2021-7. EJB acknowledges the support of CAPES under the grant PDSE 88881.690203/2022-01.  We acknowledge the utilization of the CHE cluster, which operates at the Javier Magnin Computing Center/CBPF and is funded by the COSMO/CBPF/MCTI, and received financial support from FINEP and FAPERJ. We are grateful for the availability and support of the CHE cluster, which has significantly contributed to the progress and execution of our work. 
	
	\section*{Data Availability}
	
 	The algorithms presented in this paper are openly available in the NumCosmo library at \url{https://github.com/NumCosmo/NumCosmo}.
	
	\bibliographystyle{mnras}
	\bibliography{main} 
	
	% Don't change these lines
	\bsp	% typesetting comment
	\label{lastpage}
\end{document}